\documentclass[modern]{aastex61}

\usepackage{acronym}
\usepackage{amsmath,multirow}
\usepackage{amssymb,graphicx}
\usepackage{soul}

\newcommand{\Mc}{\mathcal{M}}
\newcommand{\Msol}{\mathrm{M}_{\odot}}
\newcommand{\lten}{\log_{10}}
\newcommand{\no}{{\dot n}_0}
\newcommand{\msun}{M_\odot}

\newcommand{\Mstar}{\mathcal{M}_*}

\newcommand{\KHkl}{0.85}
\newcommand{\Skl}{0.37}
\newcommand{\Gkl}{0.39}
\newcommand{\Akl}{0.62}

\newcommand{\KHklar}{0.84}
\newcommand{\Sklar}{0.38} 
\newcommand{\Gklar}{0.39} 
\newcommand{\Aklar}{0.62}

\newcommand{\KHklbr}{2.25}
\newcommand{\Sklbr}{0.71} 
\newcommand{\Gklbr}{1.09} 
\newcommand{\Aklbr}{1.31}

\newcommand{\KHklcr}{7.11}
\newcommand{\Sklcr}{1.48} 
\newcommand{\Gklcr}{4.02} 
\newcommand{\Aklcr}{2.53}

\newcommand{\KHklb}{2.25}
\newcommand{\Sklb}{0.69} 
\newcommand{\Gklb}{1.11} 
\newcommand{\Aklb}{1.33}

\newcommand{\KHklc}{5.18}
\newcommand{\Sklc}{1.42} 
\newcommand{\Gklc}{2.86} 
\newcommand{\Aklc}{2.50}

\newcommand{\AZa}{$-1.23$}
\newcommand{\GZa}{$-1.2$}
\newcommand{\KZa}{$-2.36$}
\newcommand{\SZa}{$-0.6$}

\newcommand{\AZar}{$-1.14$}
\newcommand{\GZar}{$-1.1$}
\newcommand{\KZar}{$-2.23$}
\newcommand{\SZar}{$-0.57$}

\newcommand{\AZb}{$-2.68$}
\newcommand{\GZb}{$-3.35$}
\newcommand{\KZb}{$-5.68$}
\newcommand{\SZb}{$-1.62$}

\newcommand{\AZbr}{$-2.63$}
\newcommand{\GZbr}{$-3.17$}
\newcommand{\KZbr}{$-5.47$}
\newcommand{\SZbr}{$-1.6$}

\newcommand{\AZc}{$-5.74$}
\newcommand{\GZc}{$-8.26$}
\newcommand{\KZc}{$-13.17$}
\newcommand{\SZc}{$-3.82$}

\newcommand{\AZcr}{$-5.09$}
\newcommand{\GZcr}{$-6.38$}
\newcommand{\KZcr}{$-9.03$}
\newcommand{\SZcr}{$-3.56$}

\newcommand{\AvKa}{$1.13$}

\newcommand{\SvKa}{$1.76$}

\begin{document}

\title{
No tension between assembly models of supermassive black hole binaries and pulsar observations.
}

\author{Hannah Middleton*}

\affiliation{Institute of Gravitational Wave Astronomy and
  School of Physics and Astronomy, University of Birmingham,
  Birmingham, B15 2TT, United Kingdom}

\author{Siyuan Chen}

\affiliation{Institute of Gravitational Wave Astronomy and
  School of Physics and Astronomy, University of Birmingham,
  Birmingham, B15 2TT, United Kingdom}

\author{Walter Del Pozzo}

\affiliation{Dipartimento di Fisica ``Enrico Fermi'', Universit\`a di Pisa, Pisa I-56127 and INFN sezione di Pisa, Italy}

\author{Alberto Sesana}

\affiliation{Institute of Gravitational Wave Astronomy and
  School of Physics and Astronomy, University of Birmingham,
  Birmingham, B15 2TT, United Kingdom}

\author{Alberto Vecchio}

\affiliation{Institute of Gravitational Wave Astronomy and
  School of Physics and Astronomy, University of Birmingham,
  Birmingham, B15 2TT, United Kingdom}

\email{hannahm@star.sr.bham.ac.uk, schen@star.sr.bham.ac.uk, walter.delpozzo@unipi.it, \\ asesana@star.sr.bham.ac.uk, av@star.sr.bham.ac.uk}

\begin{abstract}

Pulsar timing arrays (PTAs) are presently the only means to search for the gravitational wave stochastic background from supermassive black hole binary populations, considered to be within the grasp of current or near future observations. 
However, the stringent upperlimit set by the Parkes PTA~\citep{ShannonEtAl_PPTAgwbg:2013, 2015Sci...349.1522S}) has been interpreted as excluding at $> 90\%$ confidence the current paradigm of binary assembly through galaxy mergers and hardening via stellar interactions, suggesting evolution is accelerated (by stars and/or gas) or stalled.
Using Bayesian hierarchical modelling, we consider implications of this upperlimit for a comprehensive range of astrophysical scenarios, without invoking stalling nor more exotic physical processes. 
We find they are fully consistent with the upperlimit, but (weak) bounds on population parameters can be inferred. 
Bayes factors between models vary between $\approx 1.03$ -- $5.81$ and Kullback-Leibler divergences between characteristic amplitude prior and posterior lie between $0.37$ -- $0.85$. 
Considering prior astrophysical information on galaxy merger rates, recent upwards revisions of the black hole-galaxy bulge mass relation~\citep{2013ARAA..51..511K} are disfavoured at $1.6\sigma$ against lighter models (eg.~\cite{2016MNRAS.460.3119S}).
We also show, if no detection is achieved once sensitivity improves by an order of magnitude, the most optimistic scenario is disfavoured at $3.9\sigma$.

\end{abstract}

\section{Implications of upper limits}
\label{sec:implications}

Dedicated timing campaigns of ultra-stable radio pulsars lasting over a decade and carried out with the best radio telescopes around the globe have targeted
the isotropic gravitational-wave (GW) background in the frequency region $\sim 10^{-9} - 10^{-7}$ Hz. No detection has been reported so far. The most stringent constraint on an isotropic background radiation has been obtained through an 11 year-long timing of 4 radio-pulsars by the Parkes Pulsar Timing Array (PPTA). It yields an upper-limit on the GW characteristic amplitude of $h_\mathrm{1 yr} = 1.0\times 10^{-15}$ (at 95\% confidence) at a frequency of 1 yr$^{-1}$ \citep{2015Sci...349.1522S}. Consistent results, although a factor $\approx 2$ less stringent, have been reported by the European PTA
(EPTA;~\cite{2015MNRAS.453.2576L}) and the North Amercian Nanohertz Observatory for Gravitational Waves (NANOGrav;~\cite{2016ApJ...821...13A}). 
The three PTA collaborations join together to form the International PTA (IPTA;~\cite{2016IPTA}).

We use the PPTA limit to place bounds on the properties of the sub-parsec population of super-massive black hole binary (SMBHBs) systems (in the mass range $\sim 10^7 - 10^{10}\, M_\odot$) in the universe and explore what constraints, if any, can be put on the salient physical processes that lead to the formation and evolution of these objects. 

We consider a comprehensive suite of astrophysical models that combine observational constraints on the SMBHB population with state of the art dynamical modelling of binary evolution. The SMBHB merger rate is anchored to observational estimates of the host galaxy merger rate by a set of SMBH-host relations \citep[][and Section~\ref{sec:models}]{Sesana:2013,2016MNRAS.463L...6S}. Rates obtained in this way are well captured by a five parameter analytical function of mass and redshift, once model parameters are restricted to the appropriate prior range (see Section~\ref{sec:models}). Individual binaries are assumed to hold a constant eccentricity so long as they evolve via three-body scattering and gradually circularize once GW emission takes over. Their dynamical evolution and emission properties are regulated by the density of the stellar environment (assumed to be a Hernquist profile \citep{1990ApJ...356..359H} with total mass determined by the SMBH mass -- galaxy bulge mass relation) and by the eccentricity during the three-body scattering phase, which we take as a free parameter. For each set of model parameters, the characteristic GW strain $h_c(f)$ at the observed frequency $f$  is computed as described in \cite{2016arXiv161200455C}, and summarised in Section~\ref{sec:models}. Our model encapsulates the significant uncertainties in the GW background due to the poorly constrained SMBHB merger rate and has the flexibility to produce a low frequency turnover due to either three-body scattering or high eccentricities. SMBHBs are assumed to merge with no significant delay after galaxies merge. As such, the models do not include the effect of stalling or delayed mergers~\citep{2016ApJ...826...11S}. 

For definiteness, we focus on the impact of the SMBH-galaxy relation by considering: an optimistic model, which we label KH13, based on \cite{2013ARAA..51..511K}, which provides a prediction of the GW background with median amplitude at $f = 1$ yr$^{-1}$ of $h_\mathrm{1yr} = 1.5\times 10^{-15}$; a conservative model (labelled G09, based on \cite{2009ApJ...698..198G}), with $h_\mathrm{1yr} = 7\times 10^{-16}$; an ultra-conservative model (labelled S16, based on \cite{2016MNRAS.460.3119S}), with $h_\mathrm{1yr} = 4\times 10^{-16}$; and finally a model that spans the whole range of predictions within our assumptions, which we label ``All''. Note that this model contains as subsets KH13, G09 and S16, but it is not limited to them. 
Details on the models are provided in Section~\ref{sec:models}.

\begin{figure}
\includegraphics[width=0.49\textwidth]{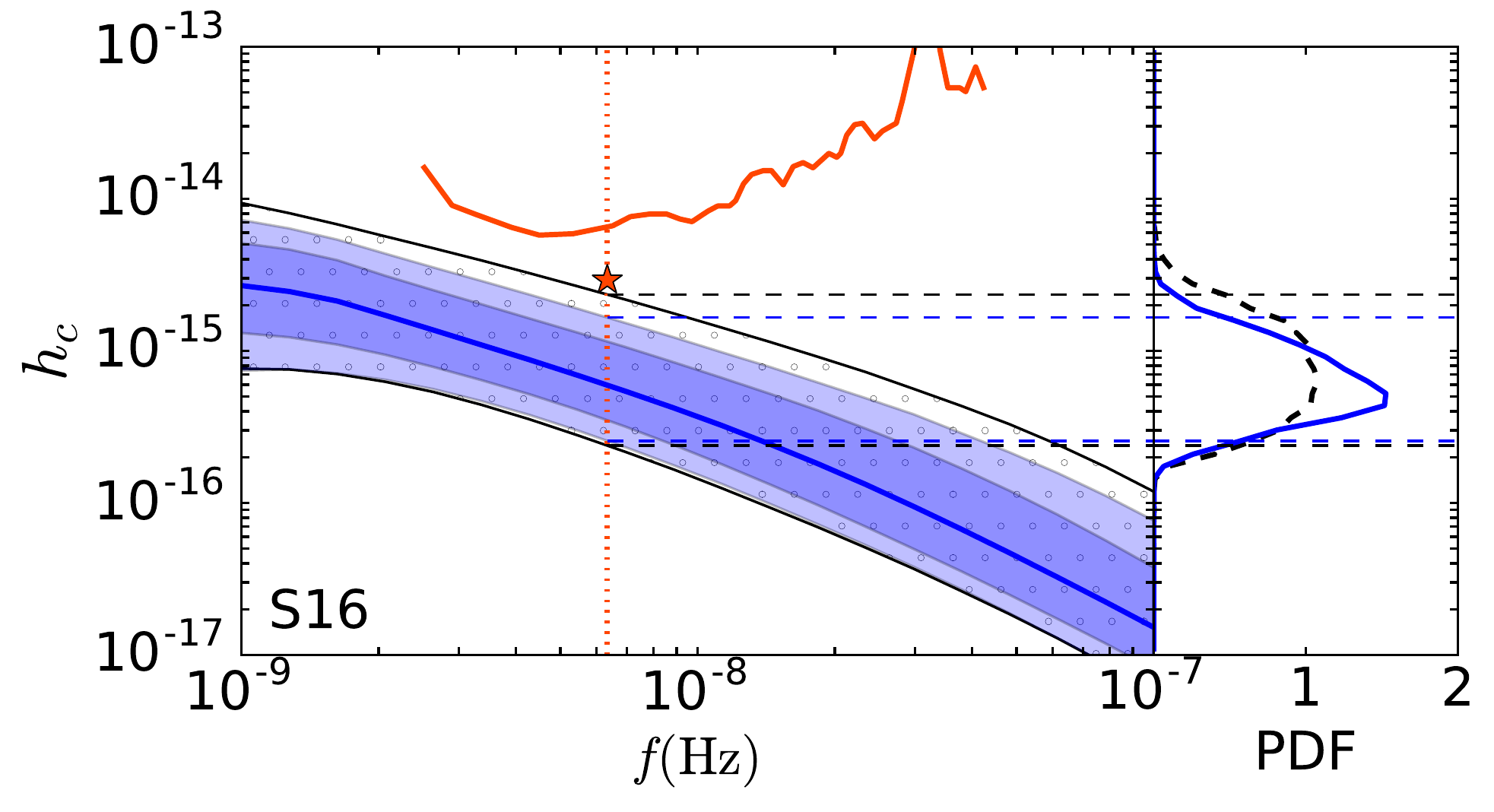}
\includegraphics[width=0.49\textwidth]{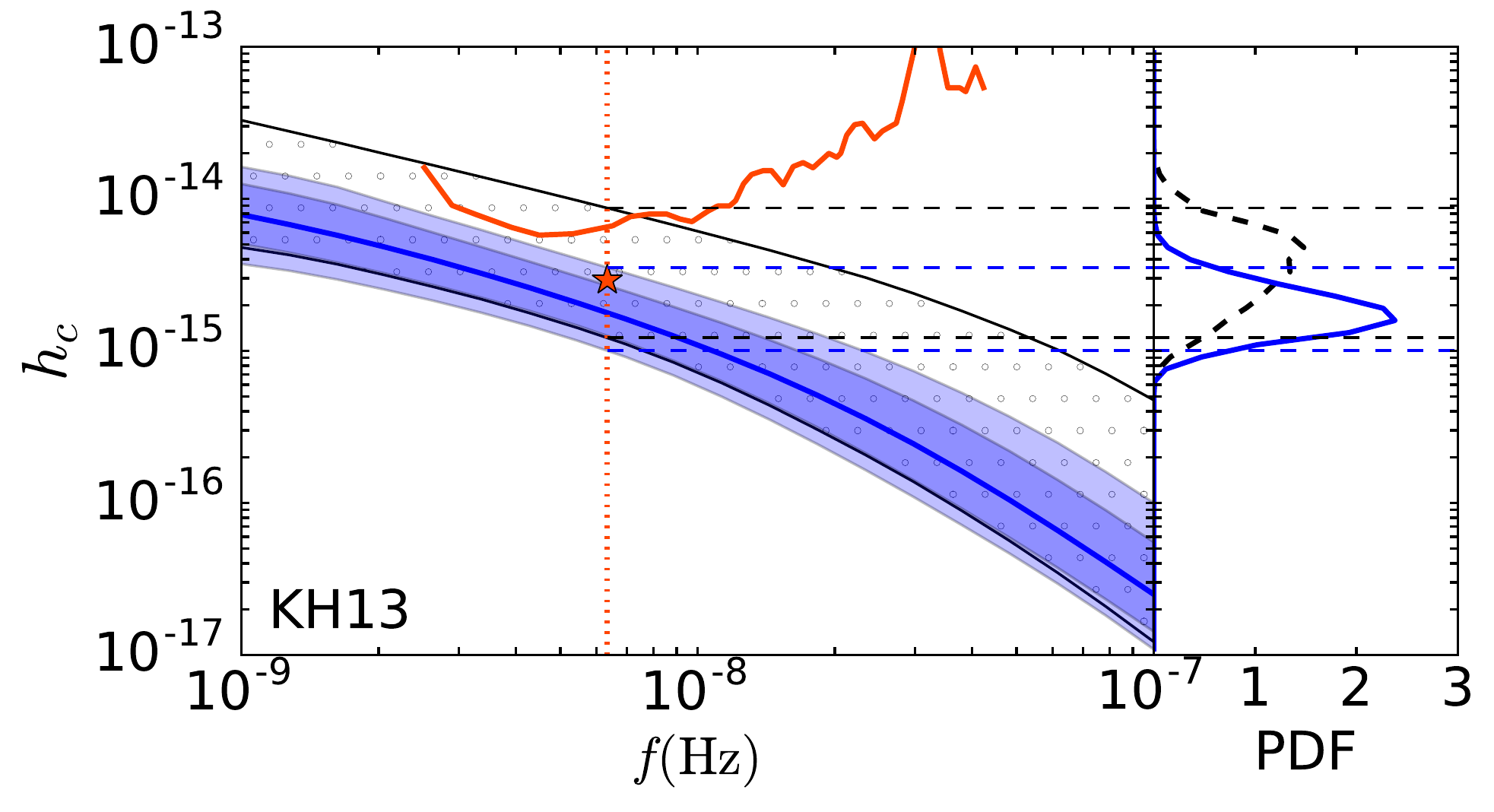}
\includegraphics[width=0.49\textwidth]{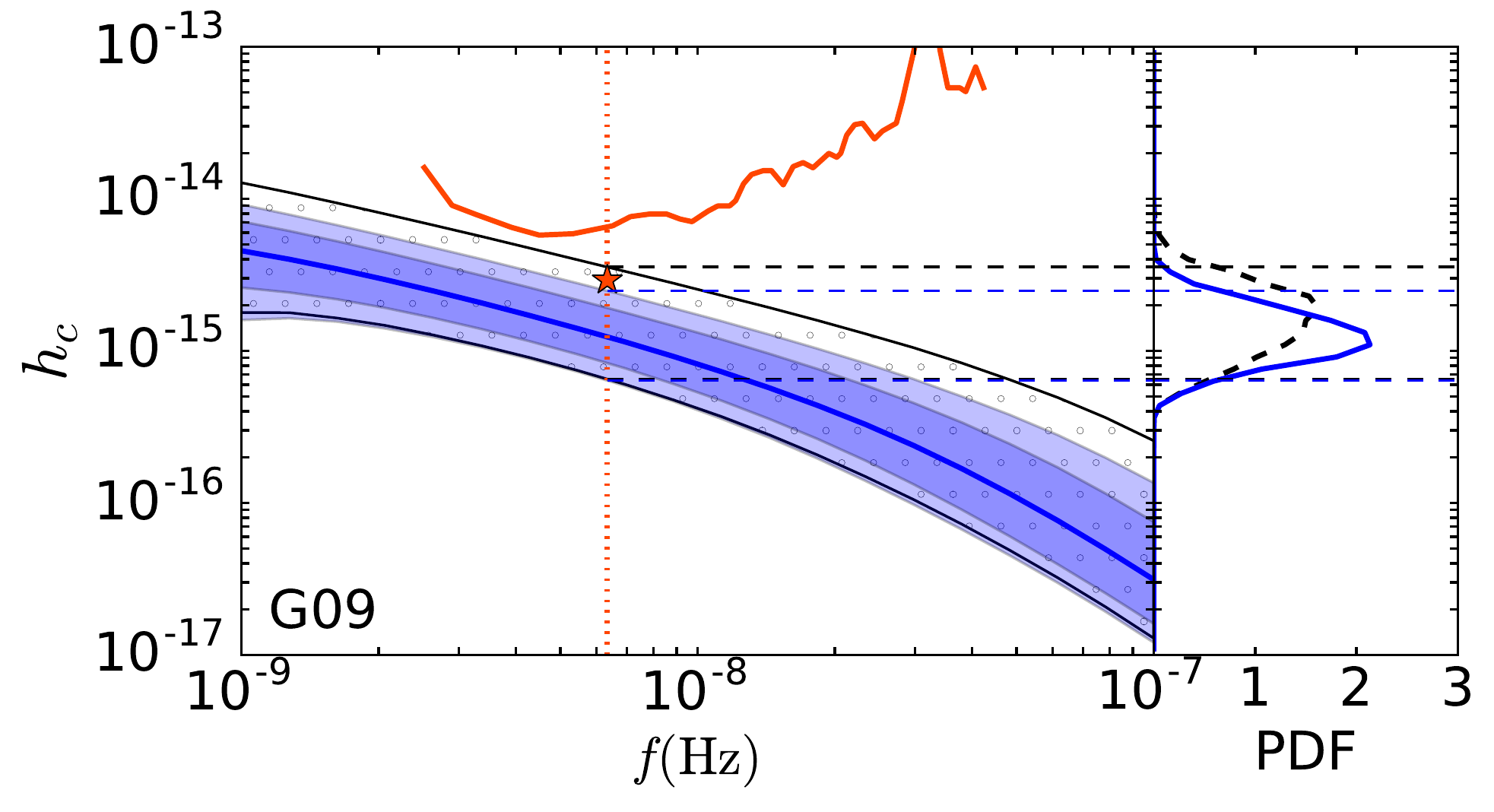}
\includegraphics[width=0.49\textwidth]{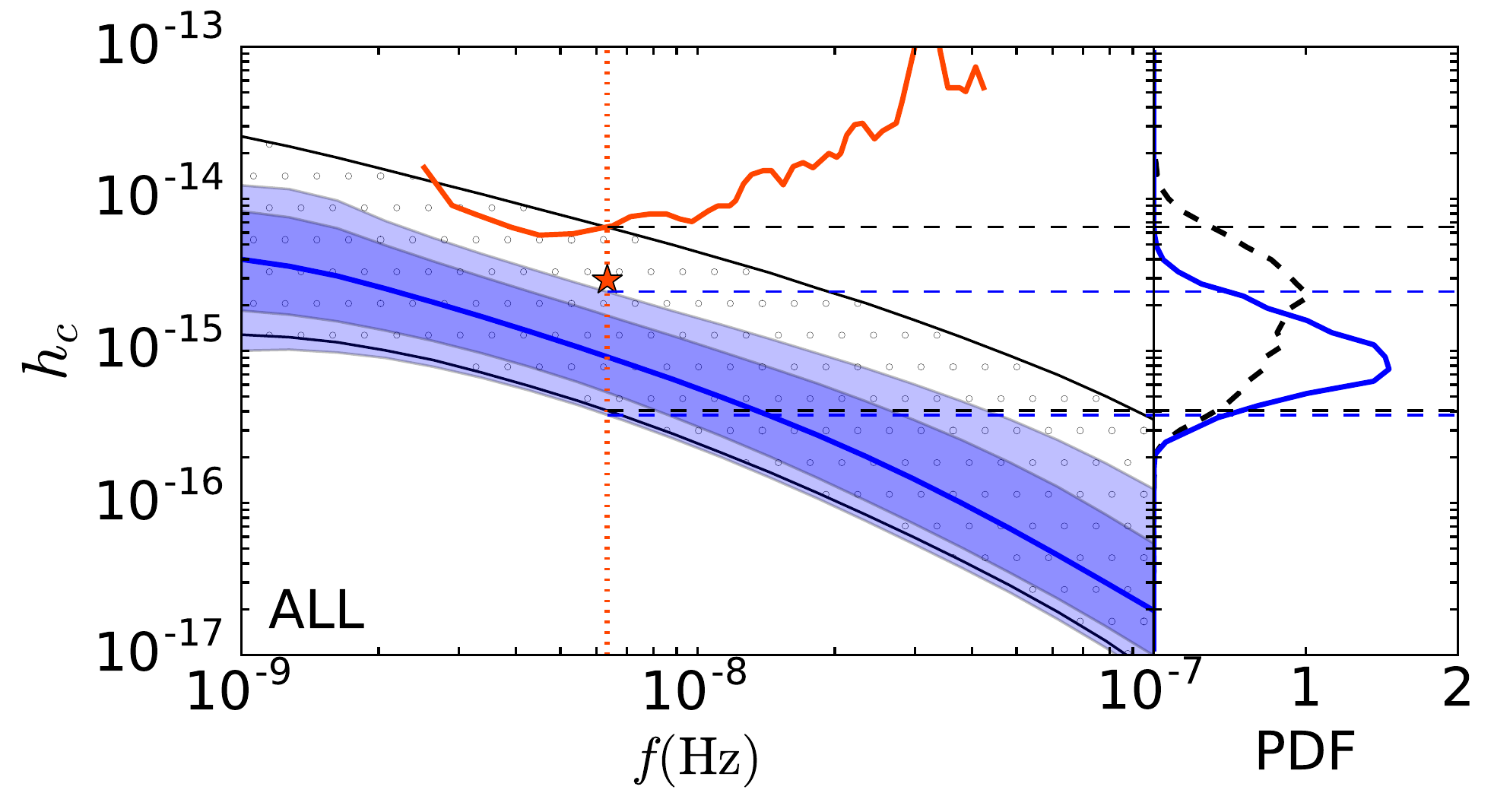}
\caption{Comparison between prior and posterior density functions on the GW stochastic background characteristic amplitude in light of the PPTA upper-limit for each of the astrophysical models considered here. The central 90\% region of the prior is indicated by the dotted band, and the posterior is shown by the progressively lighter blue shading indicating the central 68\% and 90\% regions, along with the median (solid blue line). Also shown are the PPTA bin-by-bin limit (orange solid line) and the corresponding integrated limit assuming $h_c(f)\propto f^{-2/3}$ (red star). The difference in the prior and posterior indicates how much has been learnt from the PPTA data. The right-hand side one-dimensional posterior distribution shows the prior (black-dashed) and posterior (blue-solid) at a reference frequency of $f\sim 1/5\mathrm{yr}^{-1}$, with the central 90\% regions marked (black and blue-dashed lines respectively).}
\label{fig:SpectrumPosterior}
\end{figure}

\begin{figure}
\includegraphics[width=\textwidth]{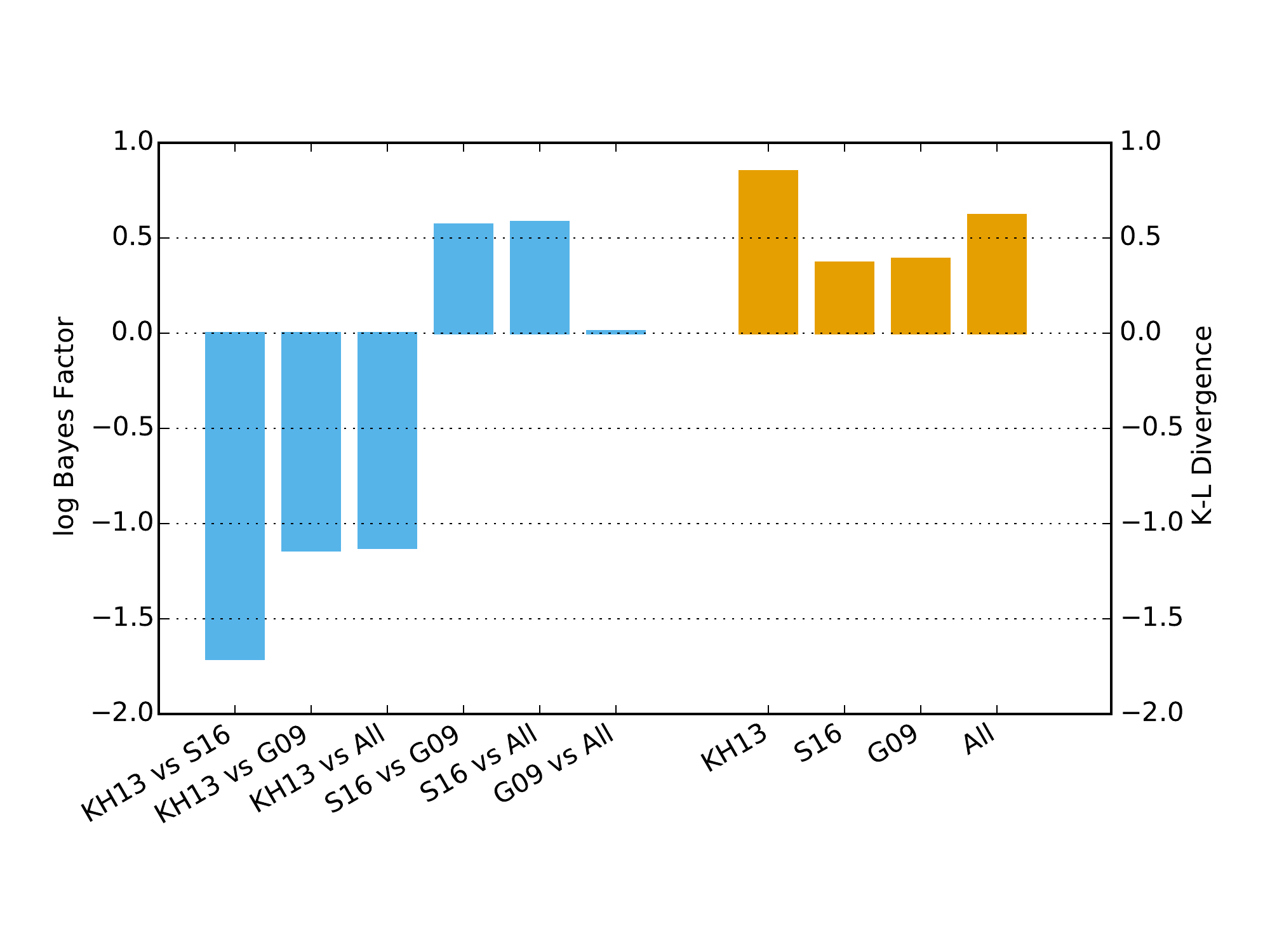}
\caption{Comparing the Bayes factors between model pairs (left hand, blue bars) and the Kullback-Leibler (K-L) divergences between the prior and posterior of characteristic amplitude (right hand, orange bars). The small range of Bayes factors, indicates that there is little to choose from between these models, although KH13 is weakly disfavoured against the others. The K-L divergences also support this conclusion. Although all values are small, KH13 has the largest K-L divergence (greatest difference between prior and posterior) of the four models. }
\label{fig:BayesFactors}
\end{figure}

For each model, we use a Bayesian hierarchical analysis to compute the model evidence  (which indicates the preference given to a model by the data and allows for the direct comparison of models) and posterior density functions on the model parameters given the data, \emph{i.e.} the posterior distribution of the GW background characteristic amplitude reported by~\cite{2015Sci...349.1522S}. We find that 
the upper limit is now beginning to probe the most optimistic predictions, but all models are so far consistent with the data. 
Figure \ref{fig:SpectrumPosterior} shows the GW characteristic strain, $h_c(f)$, of the aforementioned models. The dotted area shows the prior range 
of the GW amplitude under the model assumptions, 
and the orange solid line the 95\% confidence PPTA upper-limit on $h_c$. 
The (central) 68\% and 90\% posterior probability intervals on $h_c$ are shown by the shaded blue bands.  
The posterior density functions (PDFs) 
on the right hand side of each plot gives the prior (black-dashed line) and posterior (blue line) for $h_c$ 
at a reference frequency of $f\sim 1/5\mathrm{yr}^{-1}$. 
Figure~\ref{fig:BayesFactors} shows the natural logarithm of the ratio of the model evidence, \emph{i.e.} the Bayes factors, between all possible combinations of models and the Kullback-Leibler divergence between prior and posterior on the characteristic amplitude within a given model (with which we measure the degree of disagreement between the prior and posterior).

Qualitatively, the difference between the dotted region and the shaded bands in the main panels in Figure~ \ref{fig:SpectrumPosterior} indicates the constraining power of the Parkes PTA limit on astrophysical models -- the greater the difference between the two regions, the more suspect we are of a particular model. 
We see that although some upper portion of the allowable prior region is removed from 95\% posterior probability interval (less so for S16), none of the models can be ruled out at any significant level.
We also see that the regions covered by the confidence bands are curved (as opposed to a $h_c(f)\propto f^{-2/3}$ power-law), which one might assume to indicate the influence of the environment and eccentricity. It is important, however, to note that these are confidence bands and that although eccentricity is allowed by the data, the power-law spectrum of circular binaries driven by radiation reaction alone can clearly be consistently placed within these bands (see also Figure \ref{fig:TrianglePlots} for further details on the individual parameter posteriors including eccentricity).
This can be quantified in terms of model evidences ${\cal Z}$, shown in Table \ref{tab:KLandEvidence}. The normalization is chosen so that a putative model unaffected by the limit yields ${\cal Z} = 1$, and therefore the values can be interpreted as Bayes factors against such a model.
None of the posterior probabilities of the models with respect to this putative one show any tension, see Table \ref{tab:KLandEvidence}. For example for model All and S16 we find $e^{\textrm{\AZa}} = 0.3$ and $e^{\textrm{\SZa}} = 0.55$, respectively. Similar conclusions can be drawn from the K-L divergences, which yield $\Akl\,$ and $\Skl$. 
As a comparison, these values correspond to the K-L divergence between two Gaussian distributions with the same variance and means approximately 1.1 (for All) and 0.8 (for S16)
standard deviation apart\footnote{
The Kullback-Leibler divergence between two normal distributions $p\sim N(\mu_p, \sigma_p^2)$ and $q\sim N(\mu_q, \sigma_q^2)$ is $\mathrm{D}_\mathrm{KL}(p||q) = \ln(\sigma_q/\sigma_p) - 1/2 + 1/2 \left[(\sigma_p/\sigma_q)^2 + (\mu_p - \mu_q)^2/\sigma_q^2)\right]$. For $\sigma_p = \sigma_q$ and $\mu_p = \mu_q + \sigma_q$ the KL divergence is 0.5.
}. 
The least favourite model in the range of those considered here is KH13, with Bayes factors in favour of the others  ranging from $\approx \textrm{\AvKa}$ to $\approx \textrm{\SvKa}$. These are however values of order unity, and no decisive inference can be made from the data~\citep{kassr95}. Comparisons between each parameter's posterior and prior distribution functions are described in the supplementary material, and further support our conclusions. For KH13 -- the model that produces the strongest GW background -- we find that it has a probability of $e^{\textrm{\KZa}}=0.094$ with respect to a putative model that is unaffected by the limit. KH13 is therefore disfavoured at $\sim 1.6\sigma$. This conclusion is reflected in the value of the K-L divergence of \KHkl\footnote{This is the same K-L between two Gaussian distributions with the same variance and means approximately 1.3 standard deviation apart}. We note that \cite{2015Sci...349.1522S} choose in their analysis only a sub-sample of the~\cite{Sesana:2013} models, with properties similar to KH13. Our results for KH13 are therefore consistent with the 91\%-to-97\% `exclusion' claimed by \cite{2015Sci...349.1522S}.

\begin{table}
  \begin{center}
\begin{tabular}{|c|cc|cc|cc|}
    \hline
    \multirow{2}{*}{Model}&\multicolumn{2}{c|}{$h_\mathrm{1yr}=1\times10^{-15}$(PPTA)}&\multicolumn{2}{c|}{$h_\mathrm{1yr}=3\times10^{-16}$}&\multicolumn{2}{c|}{$h_\mathrm{1yr}=1\times10^{-16}$}\\
& K-L divergence & ${\rm log}{\cal Z}$ & K-L divergence & ${\rm log}{\cal Z}$ & K-L divergence & ${\rm log}{\cal Z}$\\
\hline
\hline
KH13   & \KHkl & \KZa & \KHklb  & \KZb & \KHklc & \KZc\\
G09    & \Gkl  & \GZa  & \Gklb  & \GZb & \Gklc  & \GZc \\
S16    & \Skl  & \SZa  & \Sklb  & \SZb & \Sklc  & \SZc \\
\hline
ALL    & \Akl  & \AZa  & \Aklb  & \AZb & \Aklc  & \AZc \\
\hline
\end{tabular}
\caption{K-L divergence and natural logarithm of the evidence ${\rm log}{\cal Z}$ for each of the four astrophysical models. Besides the PPTA upper limit at $h_\mathrm{1yr} = 10^{-15}$, we also show results for more stringent putative limits at the level of $3\times 10^{-16}$ and $1\times 10^{-16}$.}
\label{tab:KLandEvidence}
  \end{center}
\end{table}

\section{Discussion}
\cite{2015Sci...349.1522S} argue that the Parkes PTA upper-limit excludes at high confidence standard models of SMBH assembly -- \emph{i.e.} those considered in this work -- and therefore these models need to be substantially revised to accomodate either accelerated mergers via strong interaction with the environment or inefficient SMBHB formation following galaxy mergers. The work presented here does not support either claim. In particular, the posterior parameter distributions (see Section~\ref{sec:results}) favour neither high eccentricities nor particularly high stellar densities, indicating that a low frequency spectral turnover induced by SMBHB dynamics is not required to reconcile the PTA upper limit with existing models. This finding does not support an observing strategy revision in favor of higher cadence observations aimed at improving the high frequency sensitivity, as proposed by \cite{2015Sci...349.1522S}. Likewise, neither stalling nor delays between galaxy and SMBHB mergers, which, by construction, are not included in the models considered here, are needed to explain the lack of a detection of GWs at the present sensitivity level. On the other hand, PTA upper limits are now already providing interesting information about the population of merging SMBHs. The fact that KH13 is disfavoured at $1.4\,\sigma$ with respect to S16 
indicates that the population may have fewer high mass binaries, mildly favouring SMBH-host galaxy relations with lower normalizations. Although not yet decisive, our findings highlight the potential of PTAs in informing the current debate on the SMBH-host galaxy relation. Recent discoveries of over-massive black holes in brightest cluster ellipticals \citep{2011Natur.480..215M,2012MNRAS.424..224H} led to an upward revision of those relations \citep{2013ApJ...764..184M,2013ARAA..51..511K}. However, several authors attribute the high normalization of the recent SMBH-host galaxy relations to selection biases \citep{2016MNRAS.460.3119S} or to the intrinsic difficulty of resolving the SMBH fingerprint in measurements based on stellar dynamics \citep[see discussion in][]{2016arXiv160607484R}.

\section{Future prospects}

An important question is what is the sensitivity level required to really put under stress our current understanding of SMBHB assembly, if a null result persists in PTA experiments, which in turn leads to a legitimate re-thinking of the PTA observing strategy to target possibly more promising regions in the very-low frequency GW spectrum.
To address this question, we simulate future sensitivity improvements by shifting the Parkes PTA sensitivity curve down to provide 95\% upper limits of $h_\mathrm{1yr}$ at $3\times 10^{-16}$ and $1\times10^{-16}$. The results are summarised in Table \ref{tab:KLandEvidence} (more details are provided in Section~\ref{sec:results}). At $3\times 10^{-16}$, possibly within the sensitivity reach of PTAs in the next $\approx 5$ years,
S16 will be significantly favoured against KH13, with a Bayes factor of $e^{4.06}$, and only marginally over G09, with Bayes factor of $e^{1.76}$. It will still be impossible to reject this model at any reasonable significant level with respect to, say, a model which predicts negligible GW background radiation at $\sim 10^{-9} - 10^{-8}$ Hz.  However SMBH-host galaxy relations with high normalizations will show a $\approx 2\,\sigma$ tension with more conservative models. At $1\times 10^{-16}$, within reach in the next decade with the advent of MeerKAT \citep{2009arXiv0910.2935B}, FAST \citep{2011IJMPD..20..989N} and SKA \citep{2009IEEEP..97.1482D}, KH13, G09 and All are disfavoured at $3.9\,\sigma$, $2.5\,\sigma$ and $1.2\,\sigma$, respectively, with respect to S16. K-L divergences in the range $\KHklc - \Sklc$ show that the data are truly informative. S16 is also disfavoured at $2.3\sigma$ with respect to a model unaffected by the data, possibly indicating the need of additional physical processes to be included in the models.

\acknowledgements

\appendix

\section{GW background models and hierarchical analysis}
\label{sec:models}
Here we expand the description of the relevant features of our models and analysis approach.
Further details about the astrophysical models can be found in \cite{2016arXiv161200455C} and for the method see \cite{2017MNRAS.468..404C}. In Section \ref{sec:pop} we present the parametric model describing the GW background generated by a population of eccentric binaries evolving via three-body scattering. In Section \ref{sec:prior} we define the prior range of the model parameters, anchoring them to an empirical estimate of the SMBHB merger rate based on observations of close galaxy pairs. In Section \ref{sec:like} we describe the details of the implementation of Bayesian hierarchical modelling in the context of this work.

\subsection{Analytical description of the GW background}
\label{sec:pop}

The GW background from a cosmic population of SMBHBs is determined by the binary merger rate and by the dynamical properties of the systems during their inspiral. The  comoving number density of SMBHBs per unit log chirp mass ($\Mc = (M_1 M_2)^{3/5} / (M_1 + M_2)^{1/5}$) and unit redshift, $d^2 n/(d \lten \Mc d z)$, defines the normalization of the GW spectrum. If all binaries were evolving under the influence of GW backreaction only in a circular orbit, then the spectral index is also fixed at $h_c(f)\propto f^{-2/3}$ and the GW background is fully determined \citep{2001astro.ph..8028P}. To get to the point at which GW emission is efficient, however, SMBHBs need to exchange energy and angular momentum with their stellar and/or gaseous environment \citep{2013CQGra..30v4014S}, a process that can lead to an increase in the binary eccentricity \citep[e.g.][]{1996NewA....1...35Q,2009MNRAS.393.1423C}. We assume SMBHBs evolve via three-body scattering against the dense stellar background up to a transition frequency $f_t$ at which GW emission takes over. According to recent studies \citep{2015MNRAS.454L..66S,2015ApJ...810...49V}, the hardening is dictated by the density of background stars $\rho_i$ at the influence radius of the binary $r_i$. The bulge stellar density is assumed to follow a Hernquist density profile \citep{1990ApJ...356..359H} with total mass $M_*$ and scale radius $a$ determined by the SMBHB total mass $M=M_1+M_2$ via empirical relations from the literature \citep[see full details in][]{2016arXiv161200455C}. Therefore, for each individual system, $\rho_i$ is determined solely by $M$. In the stellar hardening phase, the binary is assumed to hold constant eccentricity $e_t$ up to $f_t$, beyond which it circularizes under the effect of the now dominant GW backreaction. The GW spectrum emitted by an individual binary adiabatically inspiralling under these assumptions behaves as $h_c(f)\propto f$ for $f\ll f_t$ and settles to the standard $h_c(f)\propto f^{-2/3}$ for $f\gg f_t$. The spectrum has a turnover around $f_t$ and its exact location depends on the binary eccentricity $e_t$. The observed GW spectrum is therefore uniquely determined by the binary chirp mass $\Mc$, redshift $z$, transition frequency $f_t$ and eccentricity at transition $e_t$.

The GW spectrum from the overall population can be then computed via integrating the spectrum of each individual system over the co-moving number density of merging SMBHBs:
\begin{equation}
h_c^2(f) = \int d z \int d \lten \Mc \frac{d^2 n}{d \lten \Mc d z} h^2_{c,\mathrm{fit}}\left( f \frac{f_{p,0}}{f_{p,t}} \right) \left( \frac{f_{p,t}}{f_{p,0}} \right)^{-4/3} \left( \frac{\Mc}{\Mc_0}\right)^{5/3} \left( \frac{1+z}{1+z_0}\right)^{-1/3}.
\label{eqn:hoff}
\end{equation}
$h_{c,fit}$ is an analytic fit to the GW spectrum of a reference binary with chirp mass $\Mc_0$ at redshift $z_0$ (i.e. assuming $d^2 n/(d \lten \Mc d z)=\delta(\Mc-\Mc_0)\delta(z-z_0)$), characterized by  eccentricity of $e_0$ at a reference frequency $f_0$. For these reference values, the peak frequency of the spectrum $f_{p,0}$ is computed. The contribution of a SMBHB with generic chirp mass, emission redshift, transition frequency $f_t$ and initial eccentricity $e_t$ are then simply computed by calculating the spectrum at a rescaled frequency $f(f_{p,0}/f_{p,t})$ and by shifting it with frequency mass and redshift as indicated in equation (\ref{eqn:hoff}). \cite{2016arXiv161200455C} demonstrated that this simple self-similar computation of the GW spectrum is sufficient to describe the expected GW signal from a population of eccentric SMBHBs driven by three-body scattering at $f>1$nHz, relevant to PTA measurement.

As stated above, the shape of the spectrum depends on $\rho_i$ and $e_t$. $\rho_i$ regulates the location of $f_t$; the denser the environment, the higher the transition frequency. SMBHBs evolving in extremely dense environments will therefore show a turnover in the GW spectrum at higher frequency. $e_t$ has a twofold effect. On the one hand, eccentric binaries emit GWs more efficiently at a given orbital frequency, thus decoupling at lower $f_t$ with respect to circular ones. On the other hand, eccentricity redistributes the emitted GW power at higher frequencies, thus pushing the spectral turnover at high frequencies. In our default model, $\rho_i$ is fixed by the SMBHB total mass $M$ and we make the simplifying assumption that all systems have the same $e_t$. We also considered an extended model where $\rho_i$ is multiplied by a free parameter $\eta$. This corresponds to a simple rescaling of the central stellar density, relaxing the strict $M-\rho_i$ relation imposed by our default model. We stress here that including this parameter in our main analysis yielded quantitatively identical results.

We use a generic simple model for the cosmic merger rate density of SMBHBs based on an overall amplitude and two power law distributions with exponential cut-offs, 
\begin{equation}
\frac{d^2 n}{d \lten \Mc d z} = \no \left(\frac{\Mc}{10^7\Msol}\right)^{-\alpha}e^{-\Mc/\Mstar} (1+z)^{\beta}e^{-z/z_*} \frac{d t_r}{d z}
\label{eqn:model}
\end{equation}
where $dt_r / dz$ is the standard relationship between time and redshift assuming a standard $\Lambda$CDM flat Universe with cosmological constant of $H_0 = 70 \mathrm{km s^{-1} Mpc^{-1}}$. The five free parameters are: $\no$ representing the co-moving number of mergers per Mpc$^3$ per Gyr; $\alpha$ and $\Mstar$ controlling the slope and cut-off of the chirp mass distribution respectively; $\beta$ and $z_*$ regulating the equivalent properties of the redshift distribution. Equation (\ref{eqn:model}) is also used to compute the number of emitting systems per frequency resolution bin at $f>10$ nHz. The small number statistics of the most massive binaries determines a steepening of the GW spectrum at high frequencies, full details of the computation are found in \cite{SesanaVecchioColacino:2008} and \cite{2016arXiv161200455C}. The GW spectrum is therefore uniquely computed by a set of six(seven) parameters $\theta = {\no, \beta, z_*, \alpha, \Mstar, e_t (,\eta)}$.

\subsection{Anchoring the model prior to astrophysical models}
\label{sec:prior}

Although no sub-parsec SMBHBs emitting in the PTA frequency range have been unambiguously identified to date, their cosmic merger rate can be connected to the merger rate of their host galaxies. The procedure has been extensively described in \cite{Sesana:2013}, to which we refer the reader for full details. The galaxy merger rate can be estimated directly from observations via:
\begin{equation}
\frac{d^3n_G}{dzdM_Gdq}=\frac{\phi(M_G,z)}{M_G\ln{10}}\frac{{F}(z,M_G,q)}{\tau(z,M_G,q)}\frac{dt_r}{dz}.
\label{galmrate}
\end{equation}
Here, $\phi(M_G,z)=(dn/d{\rm log}M_G)_z$ is the galaxy mass function measured at redshift $z$; ${F}(M_G,q,z)=(df_p/dq)_{M_G,z}$, for every $M_G$ and $z$, denotes the fraction of galaxies paired with a companion galaxy with mass ratio between $q$ and $q+\delta{q}$; $\tau(z,M_G,q)$ is the merger timescale of the pair as a function of the relevant parameters. We construct a library of galaxy merger rates by combining four measurements of the galaxy mass function $\phi(M_G,z)$ \citep{2013A&A...556A..55I,muzzin13,tomczak14,2016MNRAS.455.4122B}, four estimates of the close pair fraction ${F}(M_G,q,z)$ \citep{bundy09,deravel09,lopez12,xu12} and two estimates of the merger timescale $\tau(z,M_G,q)$ \citep{kit08,lotz10}. 

Each merging galaxy pair is assigned SMBHs with masses drawn from 14 different SMBH-galaxy relations found in the literature (see table \ref{tabrel}). We write them in the form
\begin{equation}
{\rm log}_{10} M=a+b{\rm log}_{10}X,
\label{scalingrel}
\end{equation}
where $X=\{\sigma/200$km s$^{-1}$, $L_i/10^{11}L_{\sun}$ or $M_*/10^{11}\msun\}$, being $\sigma$ the stellar velocity dispersion of the galaxy bulge, $L_i$ its mid-infrared luminosity, and $M_*$ its bulge stellar mass. Each relation is characterized by an intrinsic scatter $\epsilon$. $a, b, \epsilon$ are listed in table \ref{tabrel}. SMBHBs are then assumed to merge in coincidence with their host galaxy (i.e. no stalling or extra delays).

\begin{table}
\begin{center}
\begin{tabular}{ccccc}
\hline
Paper & $X$ & $a$ & $b$ & $\epsilon$\\
\hline
\cite{haring04}     & $M_*$ & 8.2 & 1.12 & 0.30\\
\cite{sani11}       & $M_*$ & 8.2 & 0.79 & 0.37\\
\cite{beifiori12}    & $M_*$ & 7.84 & 0.91 & 0.46\\
\cite{2013ApJ...764..184M}    & $M_*$ & 8.46 & 1.05 & 0.34\\
\cite{graham12}      & $M_*$ & 8.56 & 1.01 & 0.44\\
                     &               & (8.69) & (1.98) & (0.57)\\
\cite{2013ARAA..51..511K}      & $M_*$ & 8.69 & 1.17 & 0.29\\
\hline
\cite{sani11}        & $L_i$         & 8.19 & 0.93 & 0.38\\
\hline
\cite{2009ApJ...698..198G}    & $\sigma$      & 8.23 & 3.96 & 0.31\\
\cite{graham11}      & $\sigma$      & 8.13 & 5.13 & 0.32\\
\cite{beifiori12}   & $\sigma$      & 7.99 & 4.42 & 0.33\\
\cite{2013ApJ...764..184M}  & $\sigma$      & 8.33 & 5.57 & 0.40\\
\cite{grahamscott12}  & $\sigma$      & 8.28 & 6.01 & 0.41\\
\cite{2013ARAA..51..511K}   & $\sigma$      & 8.5 & 4.42 & 0.28\\
\cite{2016MNRAS.460.3119S}   & $\sigma$      & 7.8 & 4.3 & 0.3\\

\hline
\end{tabular}
\end{center}
\caption{List of parameters $a$, $b$ and $\epsilon$. See text for details. \protect\cite{graham12} proposes a double power law with a break at $\bar{M}_*=7\times10\msun$, values in parenthesis refer to $M_*<\bar{M}_*$.}
\label{tabrel}
\end{table}

\begin{figure} 
\includegraphics[width=0.49\textwidth]{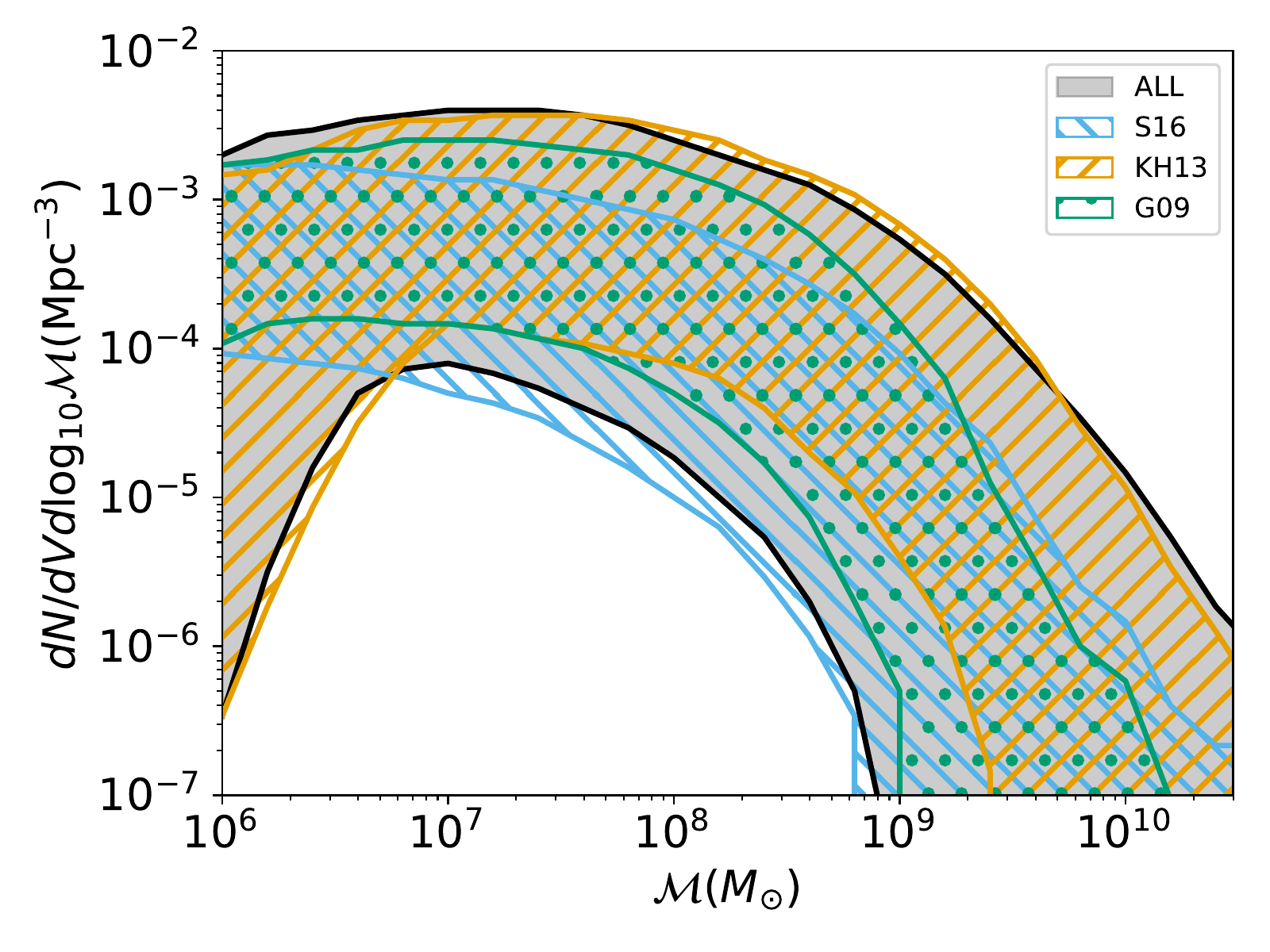}
\includegraphics[width=0.49\textwidth]{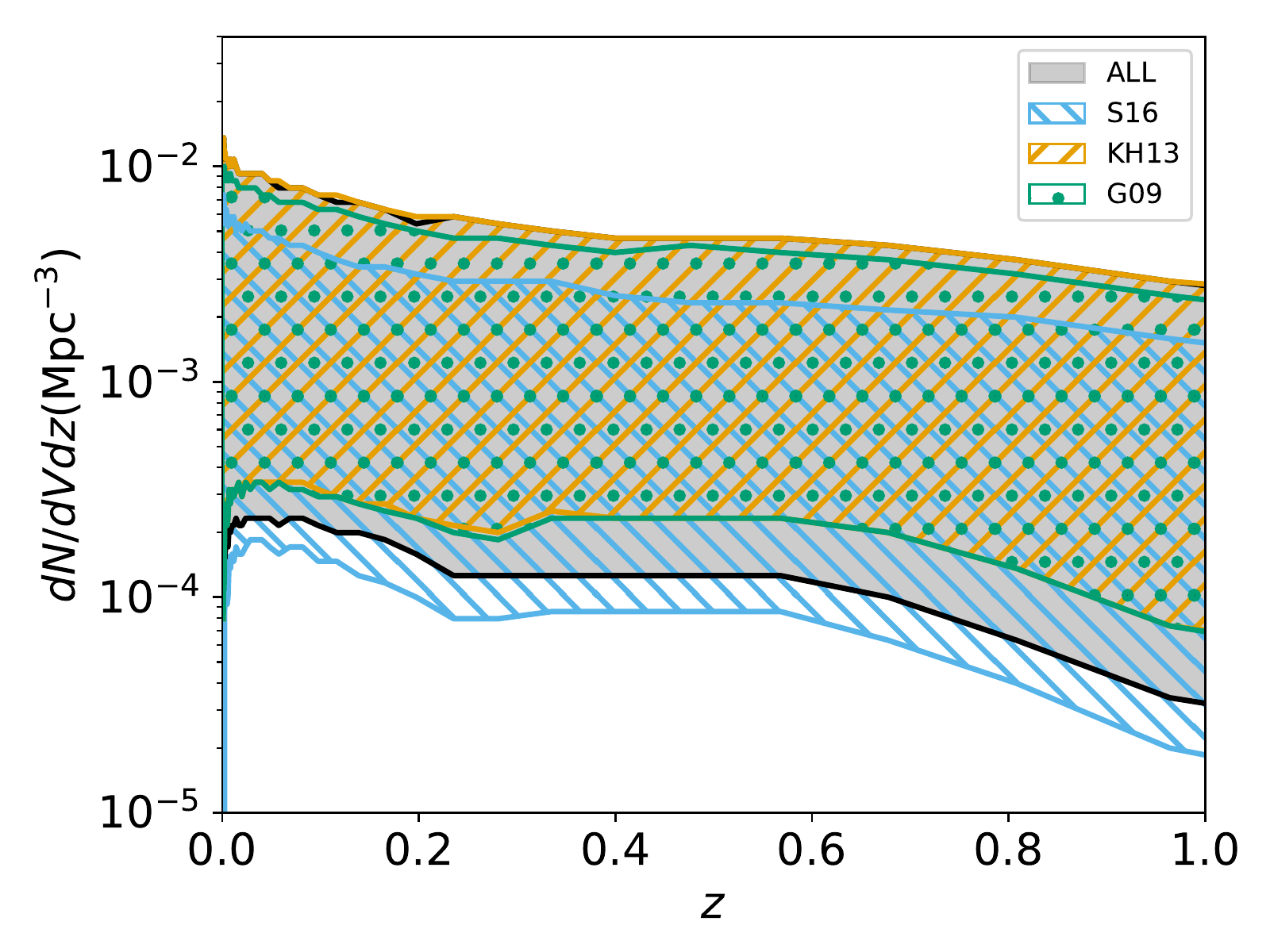}\caption{Left panel: mass density distribution $dn/d{\cal M}$ of the four astrophysical priors selected in this study (see text for full description). Right panel: redshift evolution of the SMBHB mass density for the same four models. Note that the coloured region represent the 99\% interval allowed by each model, this is why individual models can extend beyond the region associated to model All (which include KH13, G09, S16 as subsets).}
\label{fig:astroPriors}
\end{figure}

All possible combinations of galaxy merger rates as per equation (\ref{galmrate})
and SMBH masses assigned via equation (\ref{scalingrel}) result in an allowed SMBHB merger rate density as a function of chirp mass and redshift. We then marginalize over mass and redshift separately to obtain the functions $dn/dz$ and $dn/d\Mc$. We are particularly interested here in testing different SMBH-host galaxy relations, we therefore construct the function $dn/dz$ and $dn/d\Mc$ under four different assumptions: 
\begin{enumerate}
\item Model KH13 is constructed by considering both the M$-\sigma$ and M$-M_*$ relations from \cite{2013ARAA..51..511K};
\item Model G09 is based on the M$-\sigma$ relation of \cite{2009ApJ...698..198G};
\item Model S16 employs both the M$-\sigma$ relation from \cite{2016MNRAS.460.3119S};
\item Model All is the combination of all 14 SMBH mass-host galaxy relations listed in table \ref{tabrel}.
\end{enumerate}
For each of these four models, the allowed regions of $dn/dz$ and $dn/d\Mc$ are shown in figure \ref{fig:astroPriors}. The figure highlights the large uncertainty in the determination of the SMBHB merger rate and unveils the trend of the chosen models; S16 and KH13 represent the lower and upper bound to the rate, whereas G09 sits in the middle and is representative of the median value of model `All'.

\begin{figure} 
\includegraphics[width=\textwidth]{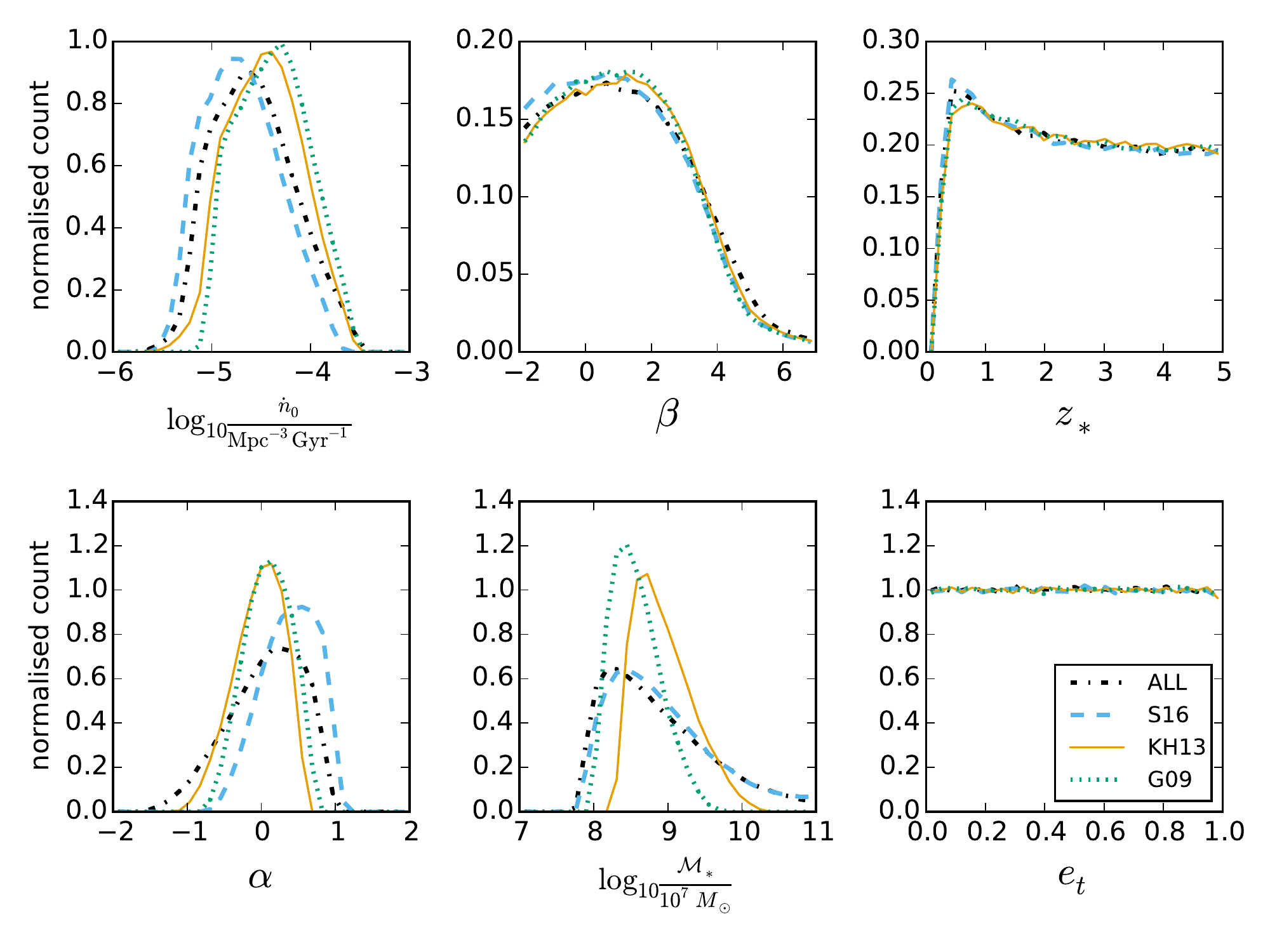}
\caption{Prior distributions of the astrophysical model parameters. Panels show: top row from left to right, $\no$, $\beta$, $z_*$; bottom row from left to right $\alpha$, $\Mstar$, $e_t$. The lines represent the prior of the four astrophysical models: KH13 (orange, solid), S16 (blue, dashed), G09 (green dotted) and ALL (black dash-dot).}
\label{fig:comparingPriors}
\end{figure}

The numerical SMBHB mass functions obtained in this way have to be described analytically by the expression (\ref{eqn:model}). Our strategy is therefore to make a large series of random draws of the five parameters defining equation (\ref{eqn:model}), and to retain only those sets that produce  $dn/dz$ and $dn/d\Mc$ within the boundaries set by the empirical models shown in figure \ref{fig:astroPriors}. The prior distributions obtained in this way are shown in figure \ref{fig:comparingPriors} for the four models. Redshift parameters ($\beta$ and $z_*$) have very similar prior for each of the models. The main differences are in the number rate density of mergers $\no$ and in the mass distribution parameters ($\alpha$ and $\Mstar$). KH13 and All prefer higher values of $\no$. On the other hand S16 allows for slightly higher values of $\alpha$ (in comparison to KH13 and G09), corresponding to a more negative slope on the mass distribution, with preference for a larger number of low mass binaries.

We then have to make sure that the distribution of characteristic amplitudes $h_c$ obtained by using the cosmic SMBHB merger rate density of equation (\ref{eqn:model}) with the prior parameters chosen as above is consistent with the $h_c$ distributions of the original models. To check this, we computed in both cases the GW background under the assumption of circular GW driven systems (i.e. $h_c \propto f^{-2/3}$) and we compared the distributions of $h_\mathrm{1yr}$, i.e. the strain amplitudes at $f=1$yr$^{-1}$. The $h_\mathrm{1yr}$ distributions obtained with the two techniques were found to follow each other quite closely with a difference of median values and 90\% confidence regions smaller than 0.1dex.
We conclude that our analytical models provide an adequate description of the observationally inferred SMBHB merger rate, and can therefore be used to constrain the properties of the cosmic SMBHB population. In particular model KH13 provides an optimistic prediction of the GW background with median amplitude at $f = 1$ yr$^{-1}$ of $h_\mathrm{1yr} \approx 1.5\times 10^{-15}$; model G09 results in a more conservative prediction $h_\mathrm{1yr} \approx 7\times 10^{-16}$; model S16 result in an ultra conservative estimate with median $h_\mathrm{1yr} \approx 4\times 10^{-16}$; and finally the characteristic amplitude predicted by the compilation of all models (All) encompasses almost two orders of magnitudes with median value  $h_\mathrm{1yr} \approx 8\times 10^{-16}$.

As for the parameters defining the binary dynamics, we assume that all binaries have the same eccentricity for which we pick a flat prior in the range $10^{-6}<e_t<0.999$. In the extended model, featuring a rescaling of the density $\rho_i$ regulating the binary hardening in the stellar phase, we assume a log flat prior for the multiplicative factor $\eta$ in the range $0.01<\eta<100$.

\subsection{Likelihood function and hierarchical modelling}
\label{sec:like}

By making use of Bayes theorem, the posterior probability distribution $p(\theta|d,M)$ of the model parameters $\theta$ inferred by the data $d$ given a model $M $ is
\begin{equation}
p(\theta|d,M) = \frac{p(d|\theta,M)p(\theta|M)}{{\cal Z}_M},
\label{eqn:BayesTheorem}
\end{equation}
where $p(\theta|M)$ is the prior knowledge of the model parameters, $p(d|\theta,M)$ is the likelihood of the data $d$ given the parameters $\theta$ and ${\cal Z}_M$ is the evidence of model $M$, computed as
\begin{equation}
{\cal Z}_M= \int p(d|\theta,M)p(\theta|M)d\theta.
\label{eqn:evidence}
\end{equation}
The evidence is the integral of the likelihood function over the multi-dimensional space defined by the model parameters $\theta$, weighted by the multivariate prior probability distribution of the parameters. When comparing two competitive models A and B, the odds ratio is computed as
\begin{equation}
{\cal O}_{A,B}=\frac{{\cal Z}_A}{{\cal Z}_B}\frac{P_A}{P_B}={\cal B}_{A,B}\frac{P_A}{P_B},  
\end{equation}
where  ${\cal B}_{A,B}={\cal Z}_A/{\cal Z}_B$ is the Bayes factor and $P_M$ is the prior probability assigned to model $M$. When comparing the four models KH13, G09, S16 and All, we assign equal prior probability to each model. Therefore, in each model pair comparison, the odds ratio reduces to the Bayes factor. In Section \ref{sec:prior} we already defined the distribution of prior parameters $p(\theta|M)$, to proceed with model comparison and parameter estimation we need to define the likelihood function $p(d|\theta,M)$.

The likelihood function, $p(d|\theta, M)$ is defined following~\cite{2017MNRAS.468..404C}. We take the posterior samples from the Parkes PTA analysis (courtesy of Shannon and collaborators) used to place the 95\% upper limit at $h_\mathrm{1yr} = 1 \times 10^{-15}$, when a single power law background $h_c\propto f^{-2/3}$ is assumed. 
However, for our analysis we would like to convert this upper limit at $f=1\mathrm{yr}^{-1}$ to a frequency dependant upper limit on the spectrum as shown by the orange curve in figure \ref{fig:SpectrumPosterior}. The likelihood is constructed by multiplying all bins together, therefore the resulting overall limit from these bin-by-bin upper-limits must be consistent with $h_\mathrm{1yr} = 1 \times 10^{-15}$. 
The $f_{\mathrm{1yr}}$ posterior distribution is well fitted by a Fermi function. To estimate a frequency dependant upper limit, we use Fermi function likelihoods at each frequency bin, which are then shifted and re-normalised in order to provide the correct overall upper limit. 
In our analysis we consider the contributions by only the first 4 frequency bins of size $1/11\,\mathrm{yr}^{-1}$, as the higher frequency portion of the spectrum provides no additional constraint. We have verified that when we include additional bins the results of the analysis are unchanged.
Ideally, we would take the bin-by-bin upper limits directly from the pulsar timing analysis to take account of the true shape of the posterior; 
however, the method we use here provides a consistent estimate for our analysis.

Having defined the population of merging binaries, the astrophysical prior and the likelihood based on the PPTA upper limit result, we use a nested sampling algorithm \citep{Skilling2004a,cpnest} to construct posterior distributions for each of the 6 model parameters. For the results shown here, we use 2000 live points and run each analysis 5 times, giving an average of around 18000 posterior samples. 
 
\section{Detailed results} 
\label{sec:results}

\begin{figure}
\centering
S16 \hspace{0.425\textwidth} KH13\\
\includegraphics[width=0.49\textwidth]{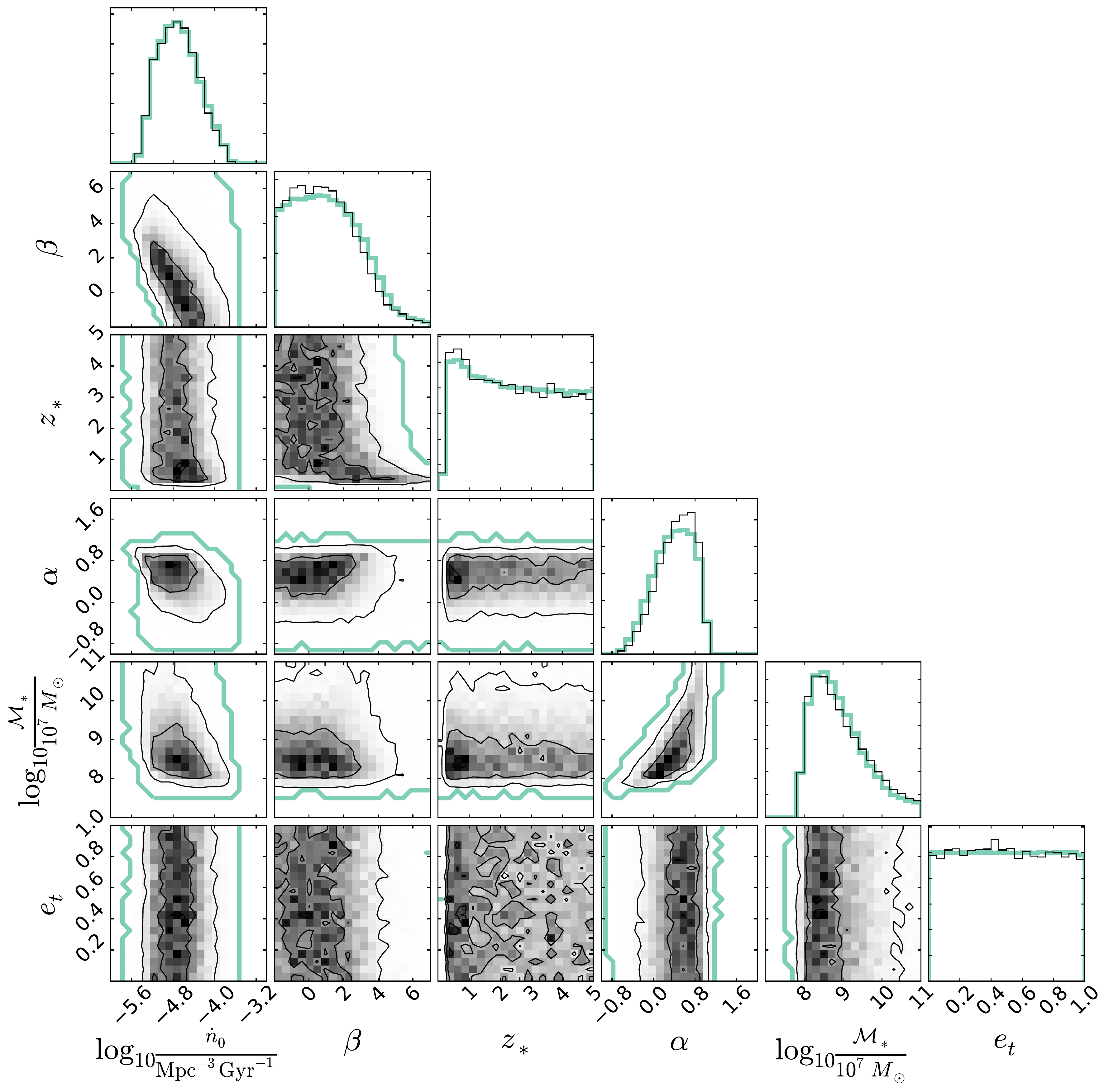}
\includegraphics[width=0.49\textwidth]{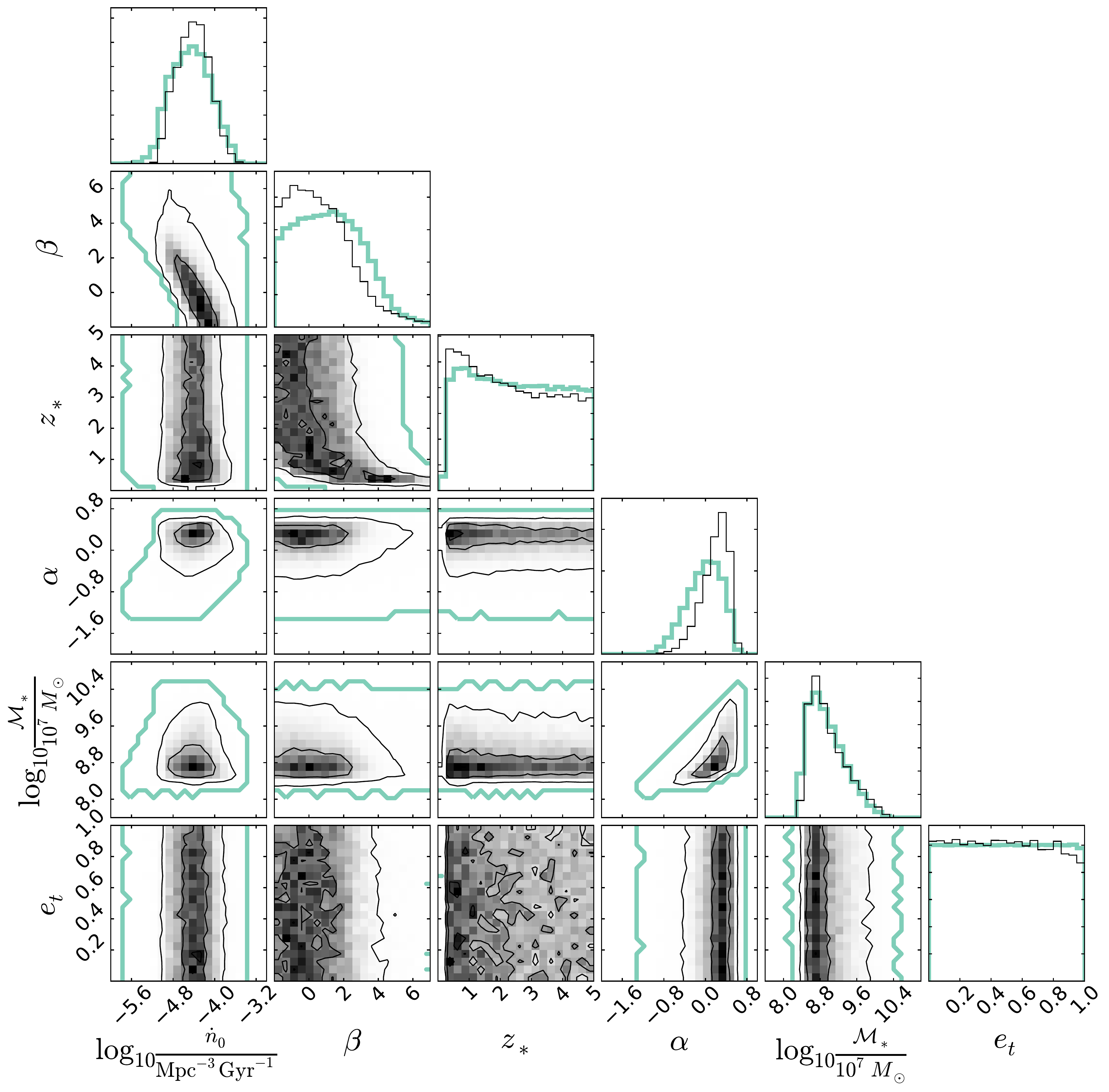}
G09 \hspace{0.425\textwidth} ALL\\
\includegraphics[width=0.49\textwidth]{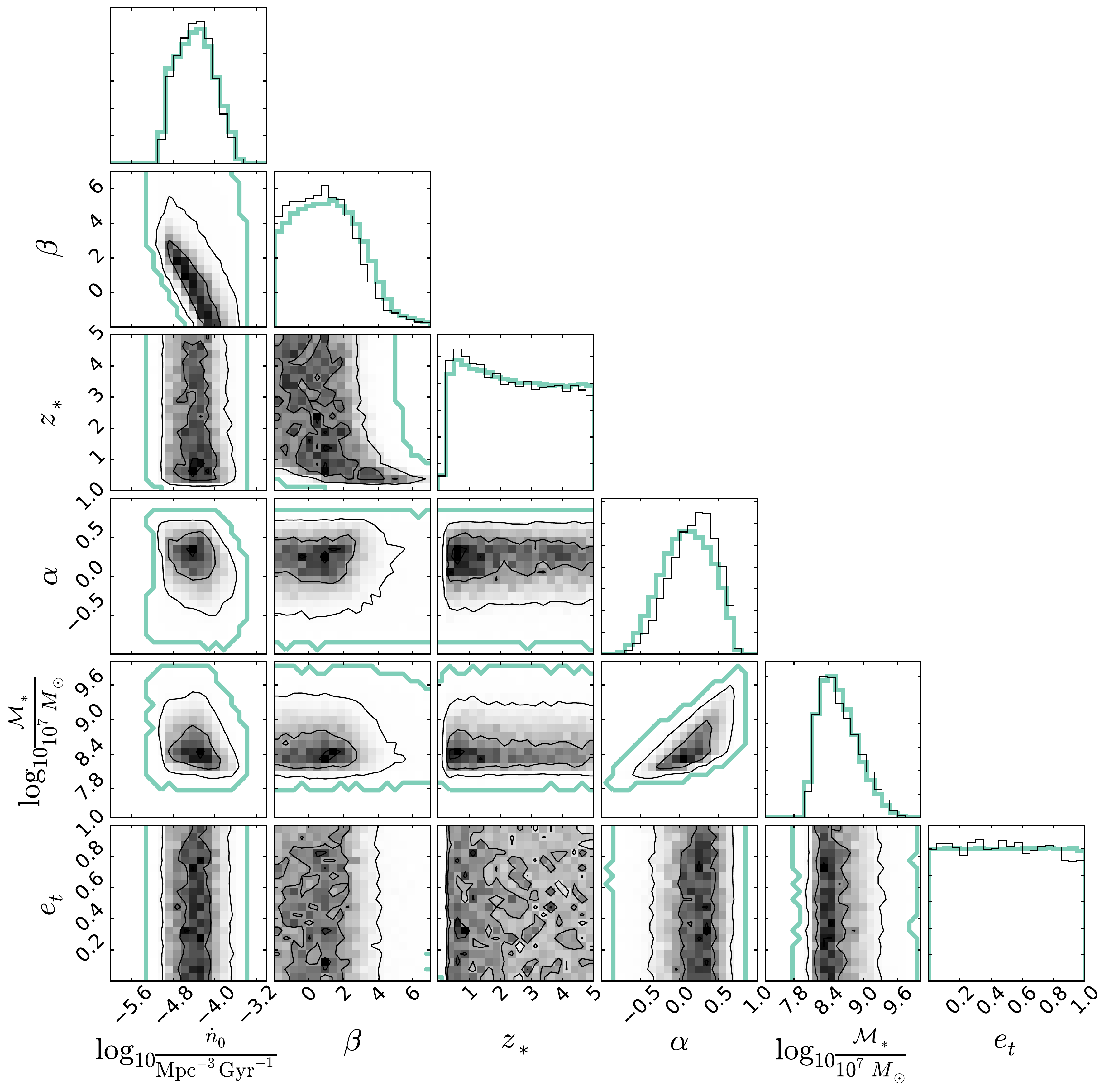}
\includegraphics[width=0.49\textwidth]{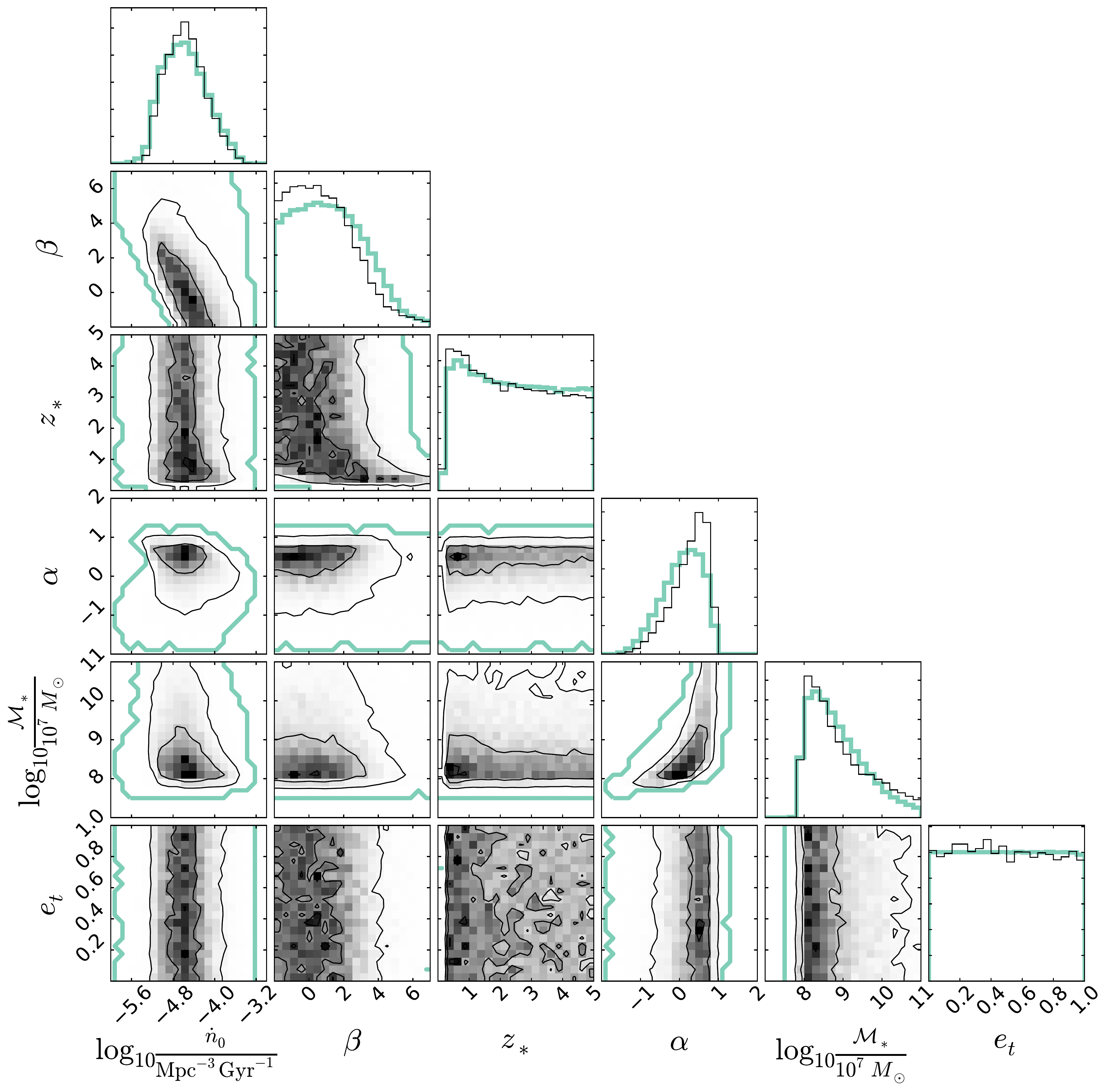}
\caption{Triangle plots for each astrophysical model showing the prior and posterior distribution for each parameter: top left S16; top right KH13, bottom left G09, bottom right All. The diagonal plots show the one-dimensional marginalised distributions for each of the 6 parameters with the thin black line indicating the posterior and the thick green line the prior. The central plots show the two-dimensional posterior distributions for each of the parameter combinations along with the green contour showing the extent of the prior.}
\label{fig:TrianglePlots}
\end{figure}
The nested sampling algorithm returns the full posterior of the N-dimensional parameter space and the value of the model evidence. The posteriors are shown in the triangle plots of figure \ref{fig:TrianglePlots} for our main analysis of the PPTA upper limit using the default six parameter model ($\theta = {\no, \beta, z_*, \alpha, \Mstar, e_t}$). 
The plots on the diagonal of the triangle show the one-dimensional marginalised distributions for each parameter whilst the two-dimensional histograms show the posterior distributions for each parameter pair.
It is immediately clear that current PTA observations impose little constraint on the shape of the SMBHB mass function. For the most conservative model (S16), the prior (green-thick lines) and posterior (black) are virtually identical (top left panel). Even for the KH13 model, the two distributions match closely, with only appreciable differences for $\beta$ and $\alpha$. This is because the PPTA limit excludes the highest values of $h_c$ predicted by the model (cf Figure \ref{fig:SpectrumPosterior}), which results in a preference for large $\alpha$ and negative $\beta$. In fact, for the mass function adopted in equation (\ref{eqn:model}), a large $\alpha$ results in a SMBHB population dominated by low mass systems, which tends to suppress the signal. Likewise, a small (or negative) $\beta$ implies a sparser population of SMBHB at higher redshift, again reducing the GW background level. In any case, little new information on the SMBHB cosmic population is acquired with current PTA measurements, which is demonstrated by the small K-L divergences between prior and posterior of the individual model parameters shown in table \ref{tab:KL}.
\begin{table}
  \begin{center}
  \begin{tabular}{|c|rrrrrr|}
    \hline
    \multirow{2}{*}{Model}&\multicolumn{6}{c|}{K-L divergence}\\
         & \multicolumn{1}{c}{$\log_{10}\dot{n}_0$} & \multicolumn{1}{c}{$\beta$} & \multicolumn{1}{c}{$z_*$} & \multicolumn{1}{c}{$\alpha$} & \multicolumn{1}{r}{$\log_{10}{\cal M}_*$} & \multicolumn{1}{c|}{$e_t$} \\
    \hline
    KH13 & $0.06$  & $0.05$  & $<0.01$ & $0.24$ & $0.03$  & $<0.01$  \\
    G09  & $<0.01$ & $0.01$  & $<0.01$ & $0.04$ & $0.01$  & $<0.01$  \\
    S16  & $<0.01$ & $<0.01$ & $<0.01$ & $0.01$ & $<0.01$ & $<0.01$  \\
    \hline
    All  & $0.02$  & $0.02$  & $<0.01$ & $0.08$ & $0.02$  & $<0.01$  \\
    \hline
  \end{tabular}
  \caption{K-L divergences of the marginalized distributions of individual parameters for the default models considered in this study as constrained by the PPTA upper limit.}
  \label{tab:KL}
  \end{center}
\end{table}

We also extended our analysis in two directions: (i) We explore a model that includes a seventh parameter, $\eta$, as described in Section \ref{sec:pop}; this parameter allows us to vary the efficiency of three-body hardening by adjusting the stellar density at the SMBHB influence radius; and (ii) We consider putative more stringent upper limits at $h_{\mathrm{1yr},95\%}=3\times10^{-16}$ and $1\times 10^{-16}$; in this case we represent this sensitivity improvement by simply lowering the 
PPTA upper limit curve by a factor of 3 and 10 respectively.

\begin{table}
  \begin{center}
  \begin{tabular}{|c|cc|cc|cc|}
    \hline
    \multirow{2}{*}{Model}&\multicolumn{2}{c|}{$h_{\mathrm{1yr},95\%}=1\times10^{-15}$}&\multicolumn{2}{|c|}{$h_{\mathrm{1yr},95\%}=3\times10^{-16}$}&\multicolumn{2}{c|}{$h_{\mathrm{1yr},95\%}=1\times10^{-16}$}\\
&$e_t$&$e_t+\eta$&$e_t$&$e_t+\eta$&$e_t$&$e_t+\eta$\\
    \hline
    KH13 &  \KZa\,(\KHkl) & \KZar\,(\KHklar)&  \KZb\,(\KHklb) & \KZbr\,(\KHklbr)&  \KZc\,(\KHklc) & \KZcr\,(\KHklcr)\\
    G09 &  \GZa\,(\Gkl) & \GZar\,(\Gklar)&  \GZb\,(\Gklb) & \GZbr\,(\Gklbr)&  \GZc\,(\Gklc) & \GZcr\,(\Gklcr)\\
    S16 &  \SZa\,(\Skl) & \SZar\,(\Sklar)&  \SZb\,(\Sklb) & \SZbr\,(\Sklbr)&  \SZc\,(\Sklc) & \SZcr\,(\Sklcr)\\
    \hline
    All &  \AZa\,(\Akl) & \AZar\,(\Aklar)&  \AZb\,(\Aklb) & \AZbr\,(\Aklbr)&  \AZc\,(\Aklc) & \AZcr\,(\Aklcr)\\
    \hline
  \end{tabular}
  \caption{Natural logarithm of model evidences and associated K-L divergences (in parenthesis) for each of the four astrophysical SMBHB coalescence rates models: KH13, G09, S16 and ALL. For each population we consider two different parametrisations of the SMBHB dynamics; one which has only $e_t$ as a free parameter (column `$e_t$' 6 parameter model), and one where we add the normalization factor $\eta$ to the density at the influence radius $\rho_i$ as a free parameter (column `$e_t+\eta$', 7 parameter model). Numbers are reported for three values of the 95\% PTA upper limit $h_{\mathrm{1yr},95\%}$, namely $10^{-15}, 3\times10^{-16}, 10^{-16}$.}
  \label{tab:all}
  \end{center}
\end{table}
The results are summarised in table \ref{tab:all}, where we list ${\rm log}{\cal Z}$ and K-L divergence (in parenthesis) of each individual model for all the performed analyses. 

Let us start by considering the implications of the current PPTA upper limit at $h_{\mathrm{1yr},95\%}= 1\times 10^{-15}$ on the extended 7-dimensional parameter models. First of all, there are no significant differences between the six and the seven parameter model. Both evidence and K-L divergence are virtually identical. Together with the flat $e_t$ posteriors shown in figure (\ref{fig:TrianglePlots}), this leads us to an important conclusion: current PTA non detections do not favour (nor require) a strong coupling with the environment. Neither high stellar densities (i.e. efficient 3-body scattering) nor high eccentricities are preferred by the data. As expected, the conservative S16 model is always favoured. However, even when compared to KH13, one obtains 
$\ln{\cal B} = 1.76$, 
which only mildly favours S16\citep{kassr95}. In addition, all K-L divergences are smaller than unity, indicating only minor updates with respect to the $h_c$ prior distributions. This is another measure of the fact that the data are not very informative.

\begin{figure*}
  \begin{minipage}[b]{0.34\textwidth}
    \begin{tabular}{c}
      \includegraphics[width=\textwidth]{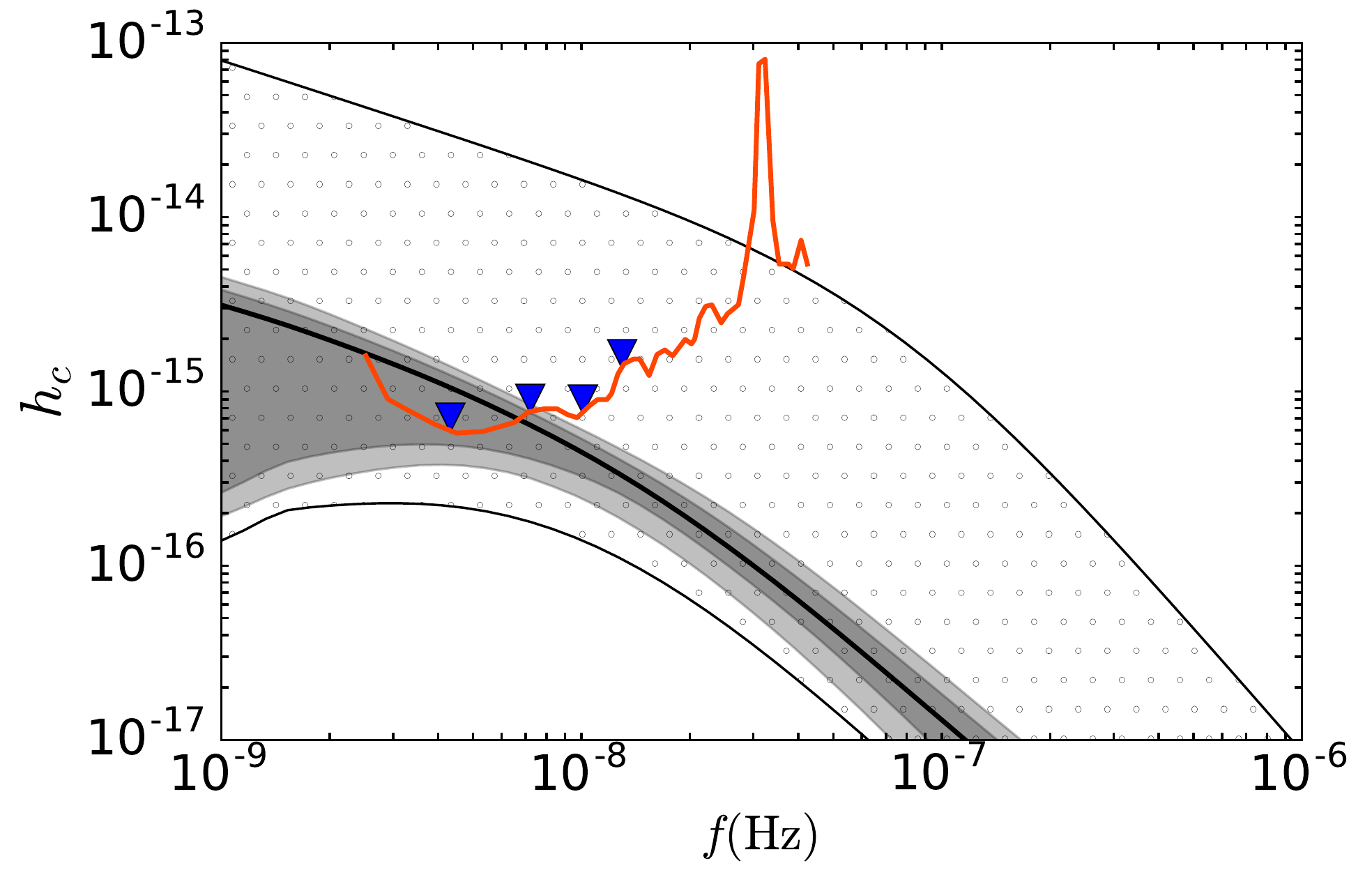}\\
      \includegraphics[width=\textwidth]{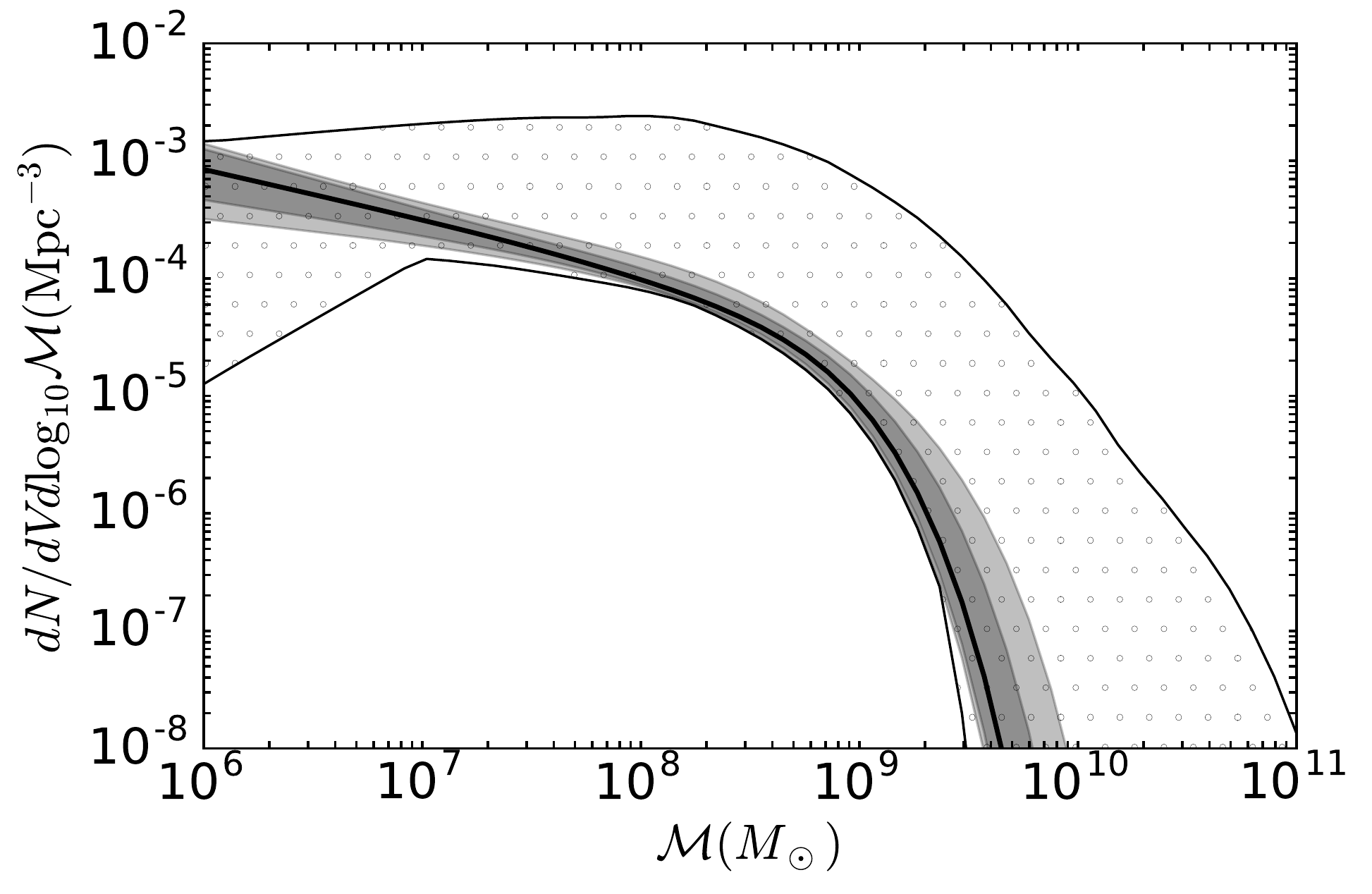}\\
      \includegraphics[width=\textwidth]{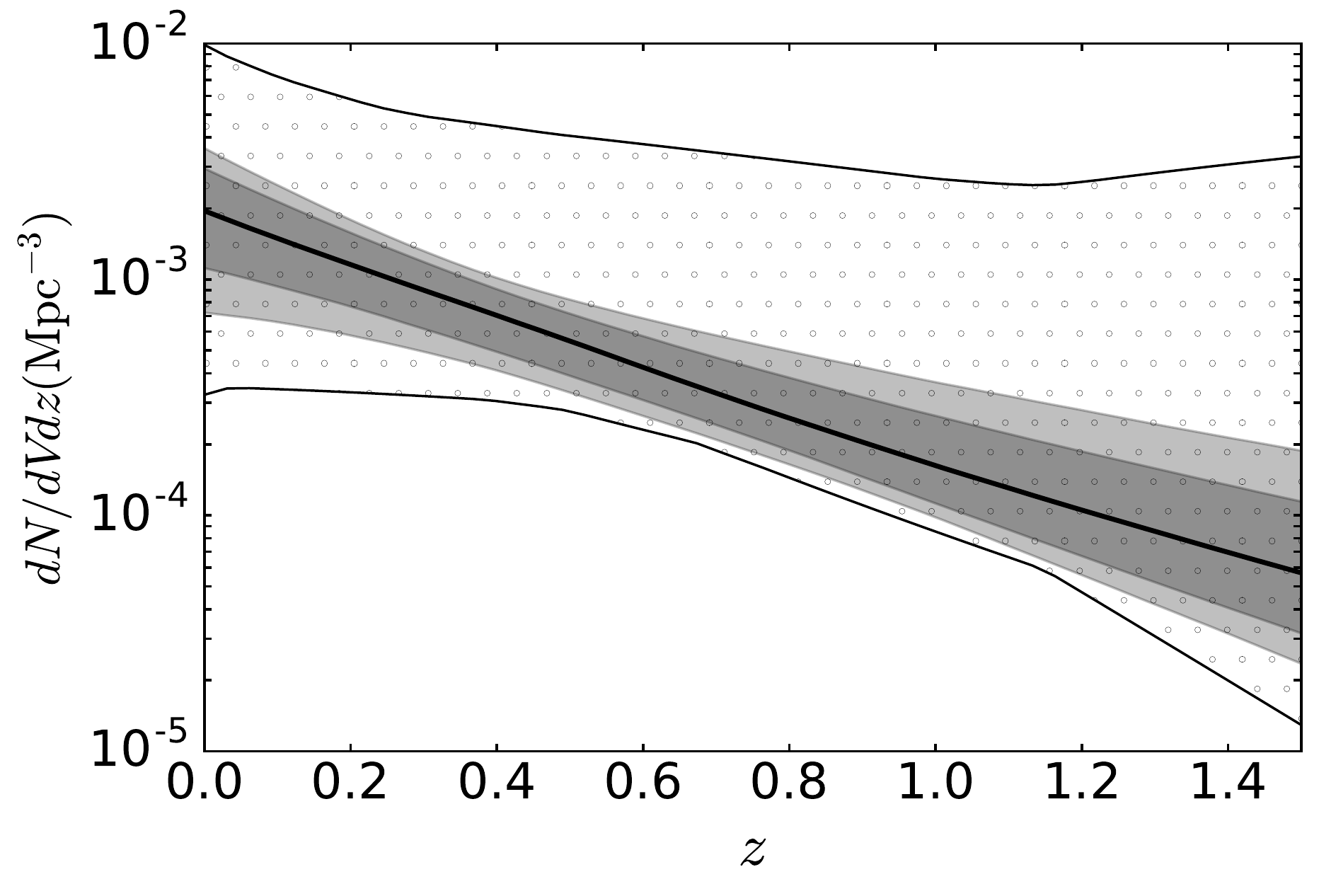}
    \end{tabular}
  \end{minipage}
  \begin{minipage}[b]{0.66\textwidth}
      \includegraphics[width=\textwidth]{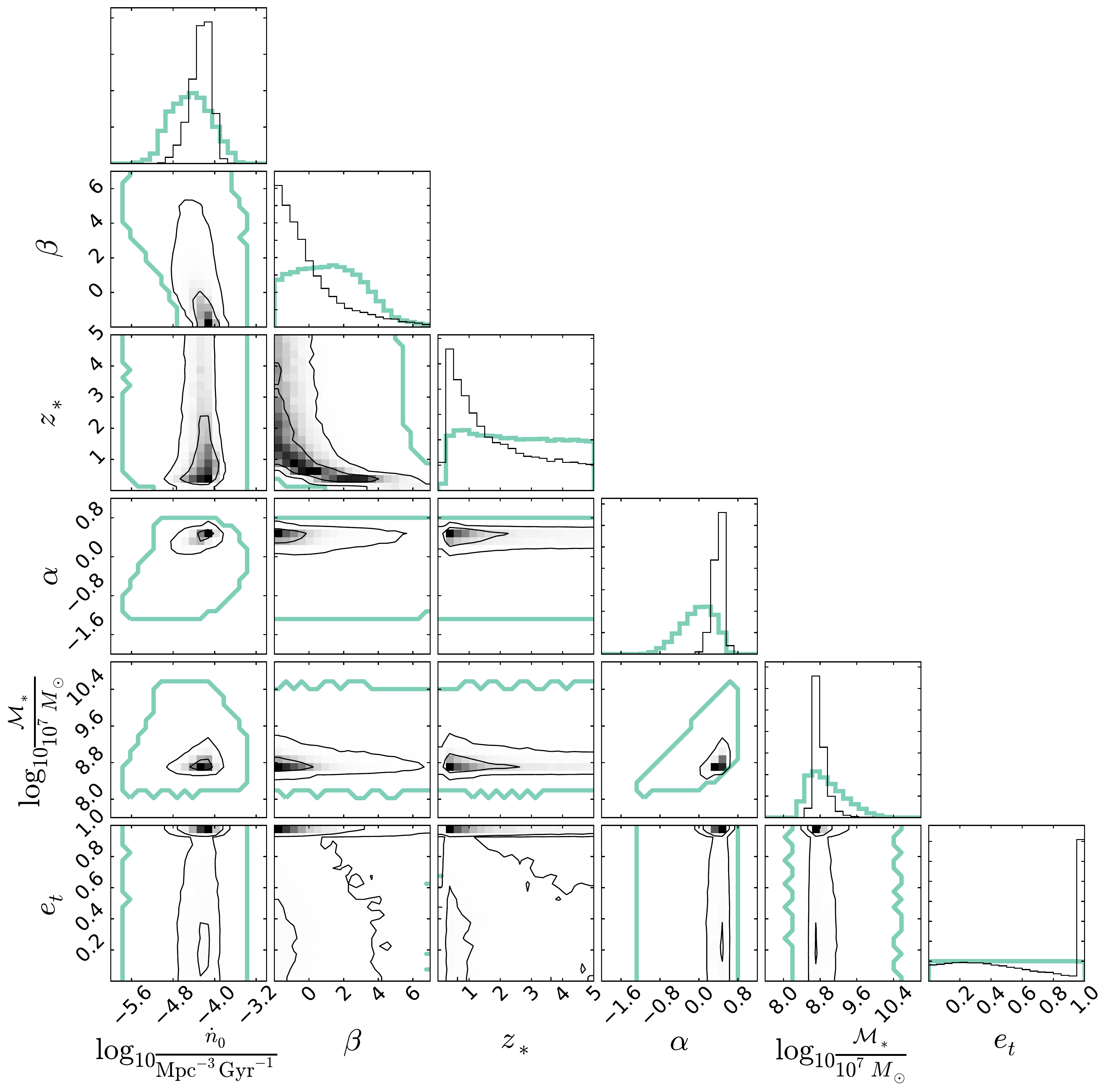}\\
  \end{minipage}
  \caption{For the six parameter model and astrophysical prior KH13. Left panel, from top to bottom: characteristic amplitude, density mass function and density redshift evolution of SMBHBs. In each panel, the dotted areas represent the astrophysical prior, the shaded bands are the 68\% and 90\% of the posterior distribution and the solid thick line is its median value. In the top panel only, the solid orange curve represents the bin-by-bin 95\% upper limits at different frequency bins (with blue triangles indicating the frequency bins we use), resulting in an overall limit $h_{\mathrm{1yr},95\%}=10^{-16}$. Right panel: the individual posterior distributions. The diagonal plots show the one-dimensional posterior distribution (black) along with the prior (green-thick), whilst the central plots show the two-dimensional posterior for each of the parameter pairs again with the extent of the prior shown by the single green-thick contour.}
  \label{fig:triangleKHO6}
  \end{figure*}

\begin{figure*}
  \begin{minipage}[b]{0.33\textwidth}
    \begin{tabular}{c}
      \includegraphics[width=\textwidth]{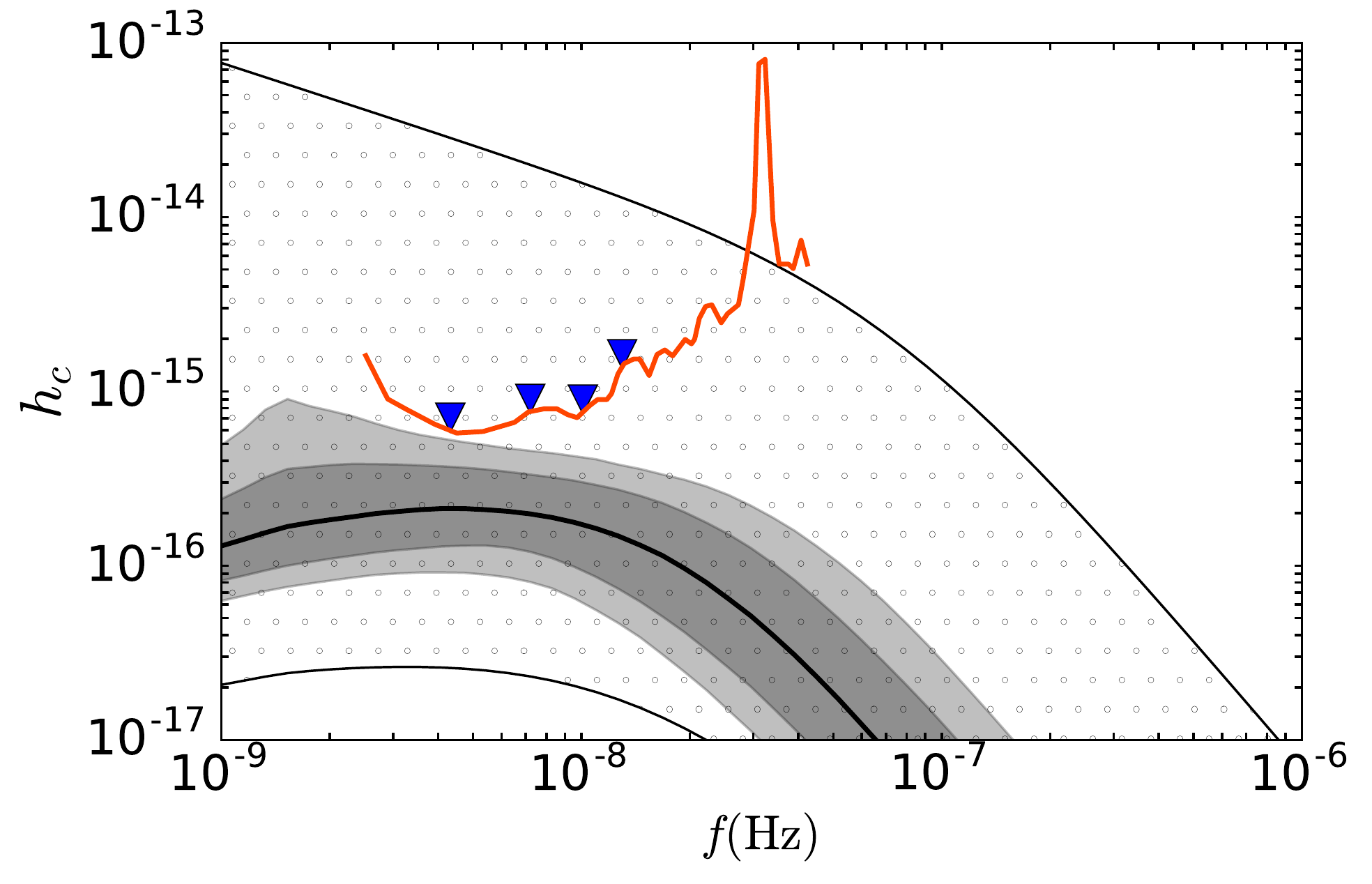}\\
      \includegraphics[width=\textwidth]{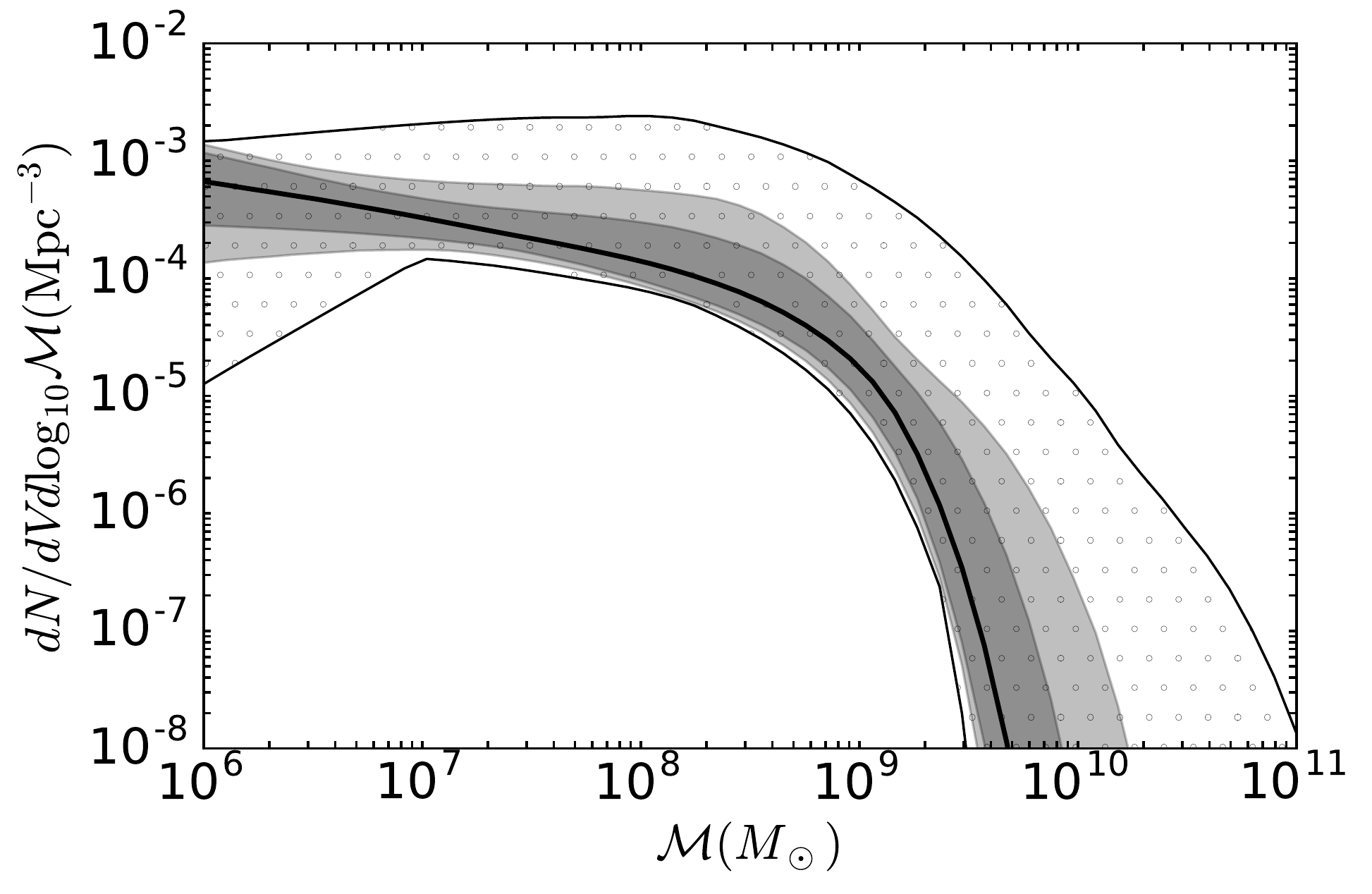}\\
      \includegraphics[width=\textwidth]{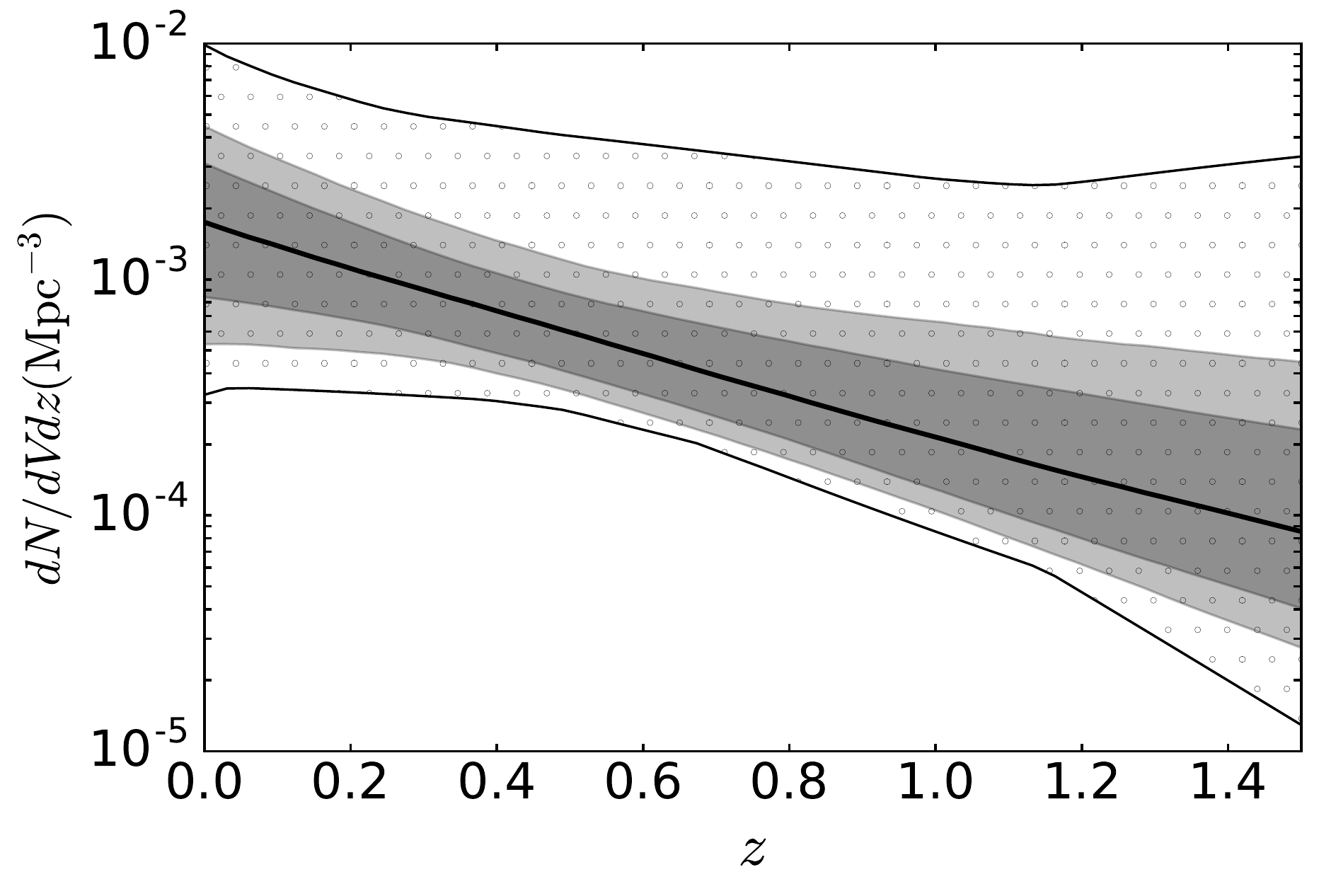}
    \end{tabular}
  \end{minipage}
  \begin{minipage}[b]{0.68\textwidth}
      \includegraphics[width=\textwidth]{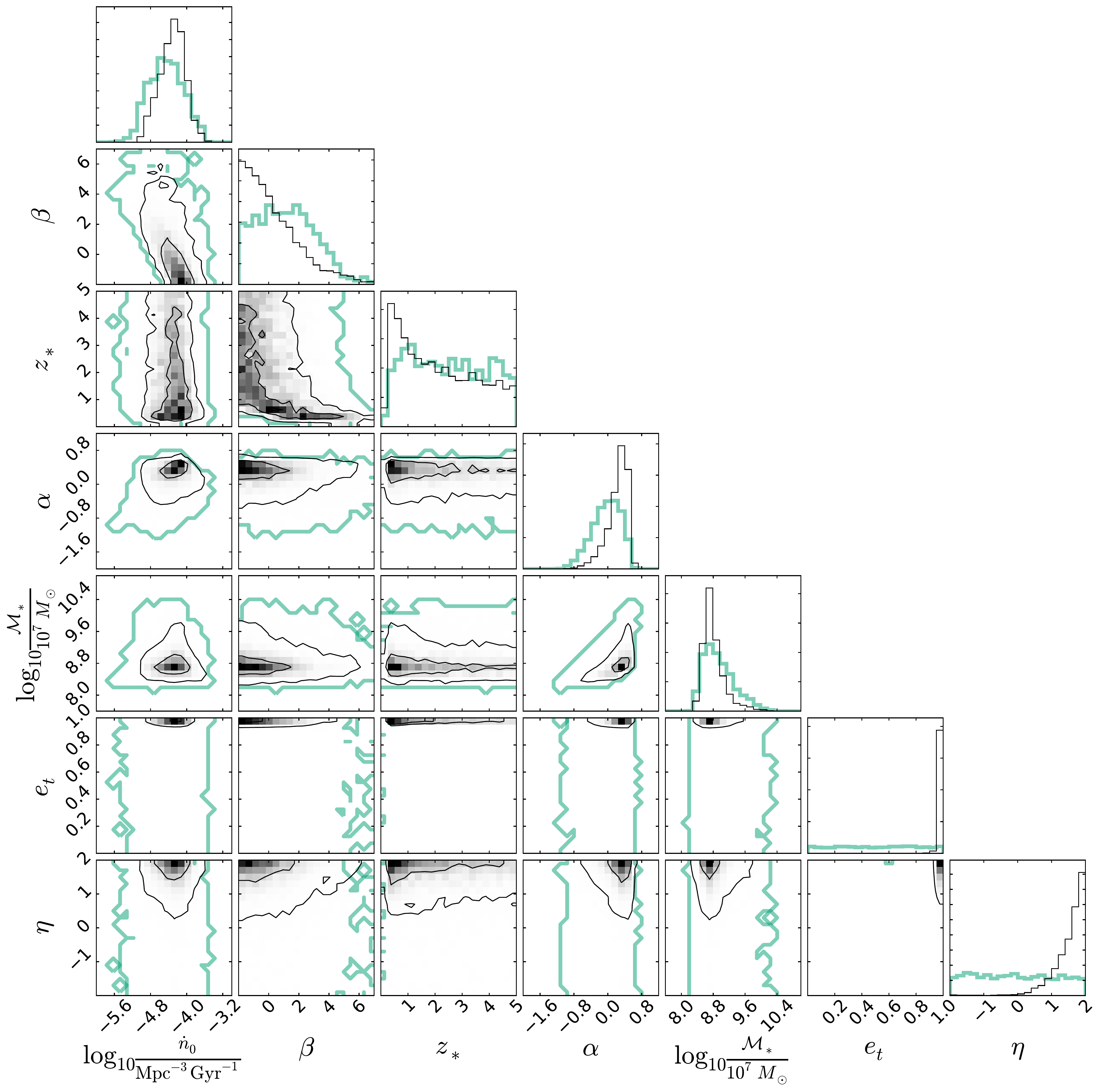}
  \end{minipage}
  \caption{Same as figure \ref{fig:triangleKHO6}, but for the seven parameter KH13 model.}
  \label{fig:triangleKHO7}
  \end{figure*}

A putative limit at $h_{\mathrm{1yr},95\%}=3\times10^{-16}$ would obviously be more constraining, as also shown by the numbers in the table. The K-L divergences of all models, with the exception of S16, are now larger than one indicating that the upper limit is becoming more informative. In terms of model comparison, S16 is now mildly favoured with respect to G09 ($\ln{\cal B} = 1.73$) 
and strongly favoured compared to KH13 ($\ln{\cal B} = 4.06$). 
We notice that again, adding $\eta$ does not make a significant difference to the model evidence. Even with such a low upper limit, neither high eccentricity nor strong coupling with the environment improve the agreement between model expectations and data. Although this seems counter-intuitive, we should keep in mind that the upper limit is set around $f\approx 5\times 10^{-9}$Hz (cf figure \ref{fig:SpectrumPosterior}). Any dynamical effect should therefore cause a turnover of the spectrum around $10^{-8}$Hz to have an impact on model selection, which occurs only in a small corner of parameter space where both $e_t$ and $\eta$ are high. However, for all models $h_{\mathrm{1yr},95\%}=3\times10^{-16}$ is still consistent with the tail of the $h_c$ distribution when an $f^{-2/3}$ spectrum is assumed, and invoking high $e_t$ and $\eta$ is not necessary.

The limit becomes far more interesting if it reaches $h_{\mathrm{1yr},95\%}=1\times10^{-16}$.
Now all K-L divergences are substantial, indicating that the measurement is indeed informative. Model selection now strongly favours model S16 compared to any other model, whether $\eta$ is included or not. Even including all environmental effects, when comparing S16 to KH13, we find $\ln{\cal B} = 5.47$,
providing decisive preference for model S16. Note however, that S16 has a log evidence of $-3.56$ of its own. This is considerably lower than zero (the evidence of a model that is unaffected by the measurement). Since delays and stalling can potentially decrease the GW background by preventing many SMBHB from merging, it is likely that a non detection at $h_{\mathrm{1yr},95\%}=1\times10^{-16}$ will provide strong support for those dynamical effects. Those are not yet included in our modelling and we plan to explore them in future work.

We have found that, contrary to the previous cases, a $1\times 10^{-16}$ limit would provide in some case significant evidence in favour of a strong coupling with the environment. To illustrate this we consider the KH13 model, where the effect is more pronounced. In this case we get $\ln{\cal B} = 4.14$
in favour of the $e_t+\eta$ model over the $e_t$ model only. Both high eccentricities and high densities would be required to explain the non detection in the context of the KH13 model. The triangle plot in figure \ref{fig:triangleKHO6} shows the posterior distribution of the model parameters for the $e_t$ case. We see that now all the posteriors differ significantly from the respective prior. Low $\beta$ and $z_*$ are preferred, because this suppresses the total number of SMBHs at high redshifts. Note that higher values of $\no$ are preferred. Although this might be surprising, it is dictated by the shape of the prior of $dn/dz$ (shown in the lower left panel in figure \ref{fig:triangleKHO6}); in order to minimize the signal, it is more convenient to allow a negative $\beta$ at the expenses of a higher local normalization $\no$ of the merger rate. High $\alpha$ values are obviously preferred, since they imply a population dominated by low mass SMBHBs (this is evident in the middle left panel of figure  \ref{fig:triangleKHO6} showing $dn/d{\cal M}$). The $e_t$ posterior now shows a prominent peak close to the maximum $e_t=0.999$, with a long tail extending to zero. Very high eccentricities are preferred, although low values are still possible. This is because $10^{-16}$ is only a $95\%$ upper limit, therefore there is a small chance that a low eccentricity model producing a signal surpassing the $10^{-16}$ value is nonetheless accepted in the posterior. The triangle plot  in figure \ref{fig:triangleKHO7} shows how the situation changes when the $\eta$ parameter is added in the $e_t+\eta$ model. Most notably, now extremely high eccentricities and high densities are strongly favoured. This is primarily because the addition of $\eta$ extends the prior in $h_c$ (shown in the upper left panel) downwards quite below the level imposed by the upper limit. It is therefore now easier to find points in the parameter space consistent with the measurement when $e_t$ and $\eta$ are large. Should other SMBH-host galaxy relations being ruled out by independent constraints, a PTA $10^{-16}$ upper limit would provide strong evidence of surprisingly extreme dynamical conditions of SMBHBs.

\begin{itemize}
\item[]\textbf{Acknowledgements:}  HM and AV acknowledge the support by the Science and Technology Facilities Council (STFC), AS is supported by a URF of the Royal Society.
\item[]\textbf{Author contributions:} All the authors have contributed to this work. 
\item[]\textbf{Competing interests:} The authors declare that they have no competing financial interests.
\item[]\textbf{Correspondence:} Correspondence and requests for materials should be addressed to H. Middleton~(email: hannahm@star.sr.bham.ac.uk).
\item[]\textbf{Data availability:} The results of our analysis for this study are available from the corresponding author on request.
\end{itemize}

\bibliography{ulPPTA}

\end{document}